# Super-compact universal quantum logic gates with inverse-designed elements

Lu He[1*], Dongning Liu[2*], Jingxing Gao[2], Weixuan Zhang[1], Huizhen Zhang[1], Xue Feng[2], Yidong Huang[2,3], Kaiyu Cui[2], Fang Liu[2], Wei Zhang[2,3+] and Xiangdong Zhang[1$]

[1]*Key Laboratory of advanced optoelectronic quantum architecture and measurements of Ministry of Education, Beijing Key Laboratory of Nanophotonics & Ultrafine Optoelectronic Systems, School of Physics, Beijing Institute of Technology, 100081 Beijing, China.*

[2]*Frontier Science Center for Quantum Information, Beijing National Research Center for Information Science and Technology (BNRist), Electronic Engineering Department, Tsinghua University, Beijing 100084, China*

[3]*Beijing Academy of Quantum Information Sciences, Beijing 100193, China.*

**These authors contributed equally to this work. [$+]Author to whom any correspondence should be addressed: zhangxd@bit.edu.cn; zwei@tsinghua.edu.cn*

**Abstract: Integrated quantum photonic circuit is a promising platform for the realization of quantum information processing in the future. To achieve the large-scale quantum photonic circuits, the applied quantum logic gates should be as small as possible for the high-density integration on chips. Here, we report the implementation of super-compact universal quantum logic gates on silicon chips by the method of inverse design. In particular, the fabricated controlled-NOT gate and Hadamard gate are both nearly a vacuum wavelength, being the smallest optical quantum gates reported up to now. We further design the quantum circuit by cascading these fundamental gates to perform arbitrary quantum processing, where the corresponding size is about several orders smaller than that of previous quantum photonic circuits. Our study paves the way for the realization of large-scale quantum photonic chips with integrated sources, and can possess important applications in the field of quantum information processes.**

## 1. Introduction

The universal quantum computer is a device capable of simulating any physical system and represents a major goal for the field of quantum information science, which can be obtained by networks of the quantum operators in universal gate sets, such as the controlled-NOT (CNOT) gate and single-qubit gates (*1*). Quantum photonic integrated circuits with CNOT gates and single-qubit gates are well recognized as attractive technology offering great promise for achieving large-scale quantum information processing (*2-9*). In recent years, many studies have been done to construct photonic quantum gates and integrated photonic circuits to perform quantum information processing (*5-7, 10-24*). At present, the footprints of silicon photonic quantum circuits to implement arbitrary two-qubit processing constructed by multi-layer Mach-Zehnder Interferometers (MZIs) are on the scale of millimeters (*4*). It is still very difficult to construct a chip in this way to perform the complexity of quantum tasks, because the number of quantum gates required increases exponentially with the increase of the quantum-state complexity to simulate an arbitrary *n* qubit quantum information process. This requires one to integrate many photonic components on an ultra-compact chip, and thus it is extremely important to reduce the size of photonic components. Until now, the pervious works report the realization of the plasmon-based (*19*) and symmetry-breaking-waveguide-based (*20*) quantum CNOT gates, whose footprints are about ~200μm$^2$ and ~21μm$^2$. Thus, how to construct optical quantum logic devices with extremely smaller sizes on the chip becomes an open problem.

On the other hand, recent investigations have shown that some inverse-designed methods can display various advantages in the design of compact optoelectronic devices (*25-43*), and many basic elements have been designed, including wavelength demultiplexers (*27*), polarization beam splitters (*28*), and so on. These inverse-designed devices possess better performances and more compact structures than those based on traditional design methods. For example, inverse-designed metastructures that solve equations have been demonstrated (*29, 30*), and an on-chip integrated laser-driven particle accelerator has been realized by the inverse design (*31*). All these studies focus on the devices in the classical electromagnetic wave systems and bring the great achievement of device miniaturization and high-performance. Additionally, the single-photon source has been theoretically designed by the inverse design method (*32*). As for the photonic quantum logic devices, inverse design methods have not been employed. It is meaningful to ask whether smaller footprints and fewer losses could appear when the inverse-designed method is applied to the design of quantum logic devices on the chip.

In this work, we design and fabricate super-compact universal quantum logic gates using the inverse-designed method on a silicon photonic chip with an integrated source.

The sizes for the fabricated optical controlled-NOT gate and single-qubit gates (Hadamard gate) on the chip are the smallest optical quantum gates ever verified in the world. Moreover, a further extension of the arbitrarily photonic quantum circuit is also provided by combining a number of CNOT and single-qubit gates.

## 2. Results

### 2.1 The inverse-designed single qubit gate.

The quantum chip is designed on the silicon-on-insulator (SOI) platform with the 220nm-thick Si layer. The schematic diagram is shown in Fig. 1a. There are four modules: (I) quantum source, (II) state preparing, (III) quantum gates, and (IV) state tomography, from the left to the right on the chip. In this work, we focus on the inverse-designed super-compact quantum gates, including the Hadamard gate, the phase z gate, and the CNOT gate, as shown in Figs. 1b-1d, respectively. The size of each area in the quantum chip can be found in S1 of Supplementary Materials.

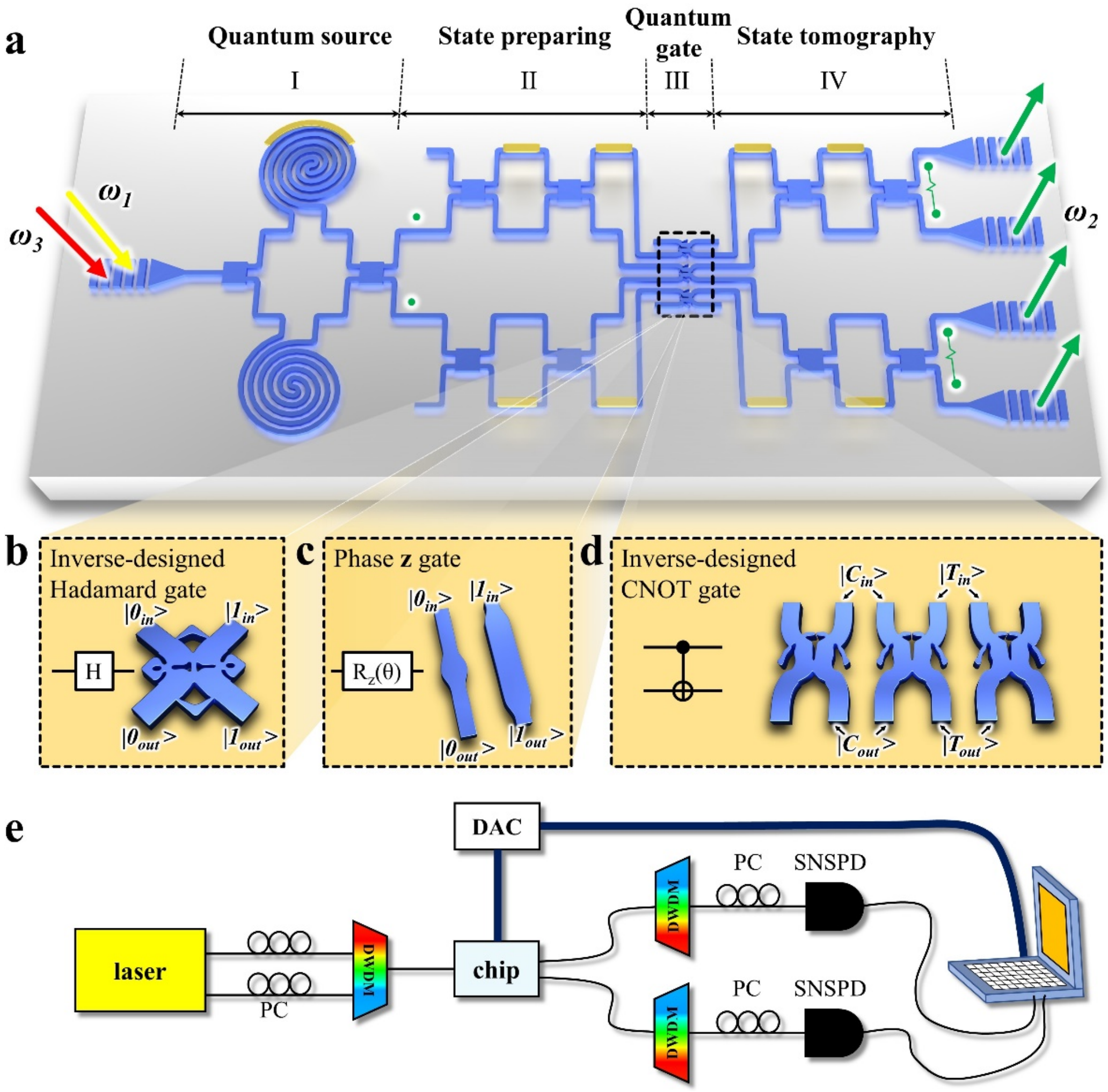

**Figure 1. The Inverse-designed super-compact quantum logic gates. a.** The schematic diagram of the photonic chip to measure the quantum gates. **b.** The inverse-designed Hadamard gate. **c.** The phase z gate. **d.** The inverse-designed CNOT gate. **e.** The schematic diagram of the experimental set-up. PC: polarization controller. DAC: digital-to-analog converter. SNSPD: superconducting nanowire single-photon detector.

Now, let us first introduce the corresponding experimental set-up to test these gates, as indicated in Fig. 1e. The two continuous-wave pump lasers of 35mW (the frequencies are $\omega_1$=194.9THz and $\omega_3$=195.3THz, and the corresponding wavelengths are 1538.19nm and 1535.04nm) are combined and injected into the experimental system. By the polarization controllers (PCs), the polarization of the pump lasers can be adjusted in order to be injected into the chip with the maximum power by the 1D transverse electric (TE) grating coupler. As shown in the left of Fig. 1a, when the pump lasers (marked as the red and yellow arrows) enter module I, it is equally split into the upper and lower waveguides by the multi-mode interference (MMI) coupler. Here, the spontaneous four-wave mixing (SFWM) process is stimulated in two 6mm-length silicon waveguides, and two photons are generated at the central frequency $\omega_2$=195.1THz ($2\omega_2=\omega_1+\omega_3$), whose corresponding wavelength is 1536.61nm. We use the thermally-tuned phase shifters (the heater made by the titanium electrode) to adjust the phase difference between two waveguides of the photon source, as marked in yellow on the Si waveguide. So, the anti-bunch state (two photons at two different waveguides) can be obtained after the second MMI. In the second module, there are two arbitrary path-encoded quantum state generators made by two MZIs and four phase shifts (PSs). Thus, the two path-encoded arbitrary single qubits can be generated by controlling four heat electrodes and injected into the inverse-designed quantum gates (the third module). The output quantum qubits are projected and detected by the fourth module of the state tomography, which is the same to the second module and made by MZIs and PSs. After going through four modules of the quantum chip, the quantum states are coupled into the fibers by 1D gratings and detected by the fiber-coupled superconducting nanowire single-photon detectors (SNSPDs). Additionally, two cascaded dense wavelength division multiplexers (cascaded DWDMs) are used to remove the residual pump photons. Nine thermal electrodes on the chip also require a computer-controlled digital-to-analog converter (DAC) for voltage controlling. Based on such a setup, the functions of designed quantum gates can be tested by analyzing the measured two-photon coincidence counts (see Methods for more details).

Here, we primarily consider the design process of the quantum gates. The first one is to design the Hadamard gate, which is one of the fundamental single-qubit gates. Generally, the 50:50 beam splitter (BS) is needed to realize the Hadamard gate on a photonic quantum chip, where the input quantum state $|\varphi\rangle$ (coming from the input port $a_{in}$) is taken as superposition state $c_1|\varphi_1\rangle + c_2|\varphi_2\rangle$. Here $c_1$ ($c_2$) represents the coefficient of the output state $|\varphi_1\rangle$ ($|\varphi_2\rangle$) from the output port $a_{out}$ ($b_{out}$). In order to carry out the optimization process, we first define an associated objective function of the BS with the single-photon excitation, $\Gamma_\lambda = c_1^2 + c_2^2$. During the design process, the objective function is maximized and the optimized structure appears. This process can be described by the following equation:

$$\max_{\varepsilon_{sio_2} \le \varepsilon(\boldsymbol{r}) \le \varepsilon_{si}} \Gamma = \sum_\lambda \Gamma_\lambda(\varepsilon(\boldsymbol{r})), \quad (1)$$

where $\varepsilon(\boldsymbol{r}) \in [\varepsilon_{sio_2}, \varepsilon_{si}]$ is the design field, it represents the material distribution of permittivities (between air and Si), λ is the wavelength, which is summed in the objective function with three different values (λ=1520nm, 1550nm, and 1580nm) to extend the range of operation frequencies. Moreover, the additional condition, $\gamma_1 \le c_1^2/c_2^2 \le \gamma_2$ ($\gamma_1$ and $\gamma_2$ are optimization parameters), should be added to limit the squares of amplitudes of output quantum states from two ports becoming nearly identical, so that the splitting ratio of 50:50 can be realized (see S2 of Supplementary Materials for detailed optimization procedures). Additionally, we also do a sensitivity analysis of our device, which ensures a good performance even if manufacturing deviations exist. We use the double filtering method (*44, 45*) in the inverse-design process. In brief, the double filtering method consists of applying the filter and threshold procedure twice on the design field, where in the second application three different threshold values are applied to obtain three different realizations of the design fields corresponding to under (over) etching. So, the over and under etching cases are simultaneously optimized. In such a way, we can get a robust device against the manufacturing defect of under (over) etching, which is the most common defect in optical chip fabrication (see S3 of Supplementary Materials for details).

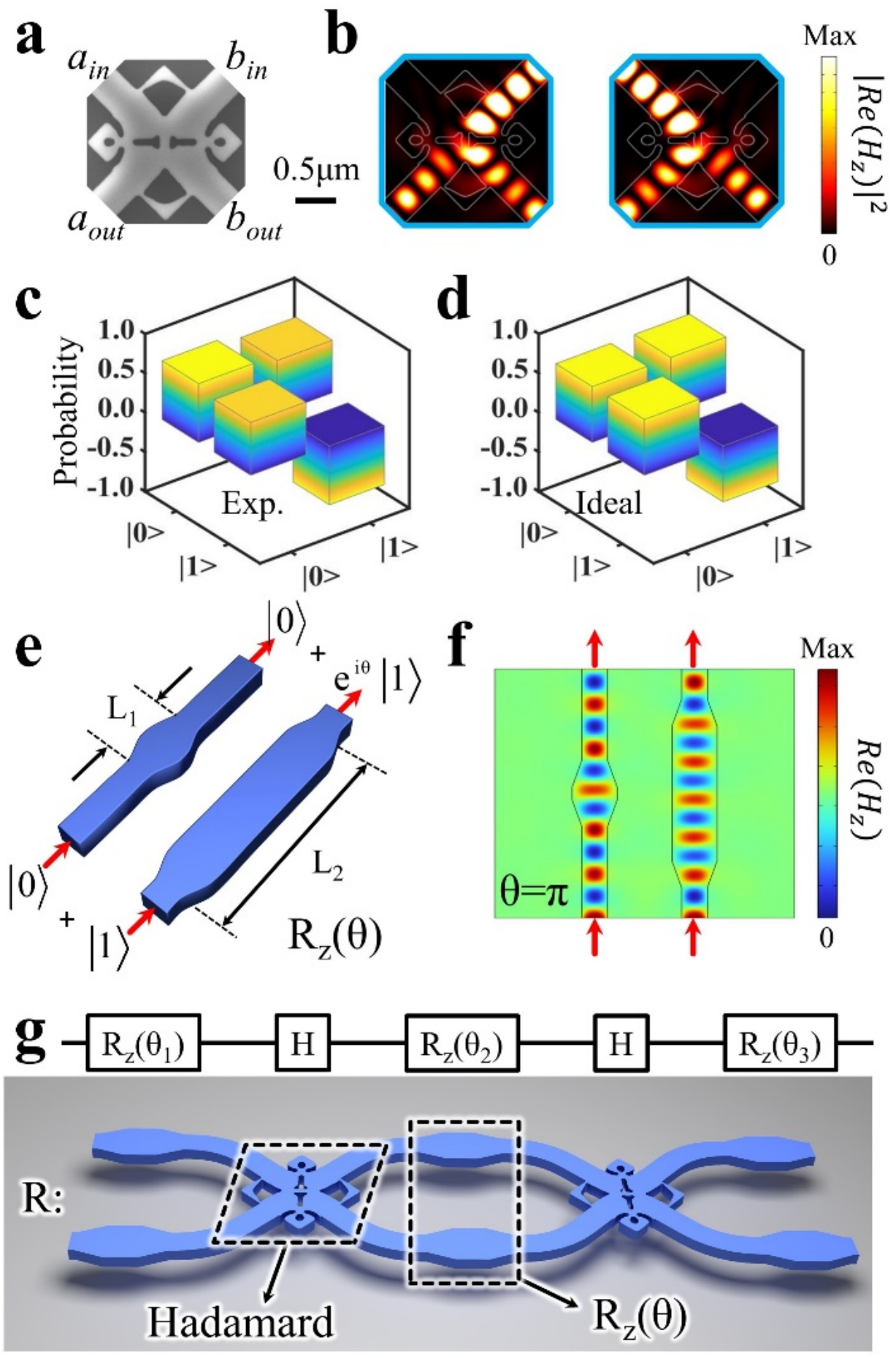

**Figure 2. The inverse-designed single-qubit gate. a.** The SEM image of the Hadamard gate. **b.** The simulation results of the 50:50 beam splitters of the Hadamard gate. **c.** The experimental matrix of the Hadamard gate. **d.** the ideal matrix of the Hadamard gate. **e.** The phase z gate $R_z(\theta)$ made by the widened waveguide. **f.** The simulation result of the phase z gate when $\theta=\pi$. **g.** The schematic diagram for the arbitrary single-qubit gate consisted of three phase z gates $R_z(\theta)$ and two Hadamard gates.

The optimized gate is fabricated using the electron-beam lithography followed by dry etching, which the SEM image is shown in Fig. 2a. The detailed fabrication process is described in Methods. It contains four 400nm-width waveguides (named $a_{in}$, $b_{in}$, $a_{out}$, and $b_{out}$, respectively) and an inverse-designed structure. The footprint of the structure is only 1.69μm$^2$ (1.3μm×1.3μm), which is less than one vacuum wavelength. In contrast, the previous works report that the footprints of Hadamard gates made by the directional coupler or MMI are about $10^2$~$10^3$μm$^2$(*4, 7*), which means that the size of the present gate is shrunk 2~3 orders compared to those of previous works.

The Hadamard operation can be performed in the designed structure based on single-photon interference. When the single-photon state is injected into the waveguide $a_{in}$ or $b_{in}$, the output photon states become superposition states with a phase difference

π/2 at $a_{out}$ and $b_{out}$. Fig. 2b shows the simulated results of field distributions under the single-photon exciting at $a_{in}$ or $b_{in}$, which indicates that the inverse-designed structure possesses the nice performances of the 50:50 BS with low loss.

To test whether such a structure can carry out the function of the Hadamard gate, we implement the single-qubit tomography of the gate. In the experiment, the input states of $|0\rangle$, $|1\rangle$, $\frac{1}{\sqrt{2}}(|0\rangle+|1\rangle)$, $\frac{1}{\sqrt{2}}(|0\rangle-|1\rangle)$, $\frac{1}{\sqrt{2}}(|0\rangle+i|1\rangle)$, and $\frac{1}{\sqrt{2}}(|0\rangle-i|1\rangle)$ are generated and injected into the Hadamard gate, where $|0\rangle$ and $|1\rangle$ represent the quantum states in the waveguide $a_{in}$ ($a_{out}$) and $b_{in}$ ($b_{out}$), respectively. After going through the Hadamard gate, the output quantum states are projected to the six states (the same to the input states). Thus, the projection probabilities of these output states for all input states are recorded in a 6×6 matrix. Based on these measurement data, we retrieve the experimental transformation matrix with the $|0\rangle$ and $|1\rangle$ bases, as shown in Fig. 2c. It is clearly seen that indeed the input states for $|0\rangle$ and $|1\rangle$ are successfully transformed into $\frac{1}{\sqrt{2}}(|0\rangle+|1\rangle)$ and $\frac{1}{\sqrt{2}}(|0\rangle-|1\rangle)$, respectively. The experimental matrix of the Hadamard operation is very close to the ideal one, which is shown in Fig. 2d. For quantitatively characterizing the Hadamard gate (*46*), we calculate its fidelity $F_H$ being 0.987(3), which is defined as $F_H = |\langle\phi|M_{th}M_{exp}|\phi\rangle|^2$. Here, $M_{th}$ and $M_{exp}$ are the theoretical and experimental matrices of the Hadamard gate, and $|\phi\rangle$ is defined as quantum state $|0\rangle$ or $|1\rangle$. Such a high fidelity further indicates that the function of the Hadamard gate is well implemented.

Based on such a Hadamard gate, combined with a phase rotation z gate $R_z(\theta)$, we can construct an arbitrary single-qubit gate R. For the path-encoded scheme, $R_z(\theta)$ is easy to be realized by introducing the phase difference of the photon state between two paths. As shown in Fig. 2e, the phase z gate is constructed by different lengths of the widened waveguides with the width being 700nm and the lengths being $L_1$ and $L_2$ (less than 2.5μm). When $L_1 \neq L_2$, there is a phase difference θ between the quantum states $|0\rangle$ and $|1\rangle$ in these waveguides. Thus, the phase z gate can be realized. To further characterize the quality of $R_z(\theta)$, a special case (the phase θ=π) is simulated and the field distribution is shown in Fig. 2f. The simulation result indicates that the phase z gate possesses an ultra-low loss and precise phase π attaching to the quantum state $|1\rangle$.

And then, an arbitrary single-qubit gate R is constructed by combining three $R_z$ and two Hadamard gates, as shown in Fig. 2g. The phases of three $R_z$ are $\theta_1$, $\theta_2$, and $\theta_3$ respectively, which can be adjusted to the fixed values to map the single-qubit state to any point on the Bloch Sphere. By the simulation (see S4 of Supplementary Materials for details), a high performance of the single-qubit gate R is also demonstrated.

**2.2 The inverse-designed CNOT gate.**

Now, we inverse-design the two-qubit CNOT gate by using the linear optical scheme. Such a scheme has been demonstrated in free space (*2, 10*) and integrated optics (*4-7, 17-20*). The design of the CNOT gate can be realized by combining three 33:67 BSs (the transmittance is T=0.67, and the reflectivity is R=0.33) in parallel. In principle, the target photon is flipped when the two-photon interference happens between the control and target photons. Thus, the inverse design of the CNOT gate first requires the inverse design of the 33:67 BS. The inverse design process of the 33:67 BS is similar to that of the Hadamard gate, except for changing the optimization parameters $\gamma_1$ and $\gamma_2$ in the design process (see S2 of the supplementary materials for details). The designed and fabricated CNOT gate is shown in Fig. 3a. The fabrication process of the CNOT gate is also similar to that of the Hadamard gate. The 33:67 BSs are spaced 500nm apart to ensure that the quantum states inside do not interfere with each other. The width of the CNOT gate is 6.4μm and the depth is 1.3μm (less than one vacuum wavelength) in the direction of quantum state propagating. The footprint of the designed CNOT gate (8.32μm$^2$) is the smallest in the world.

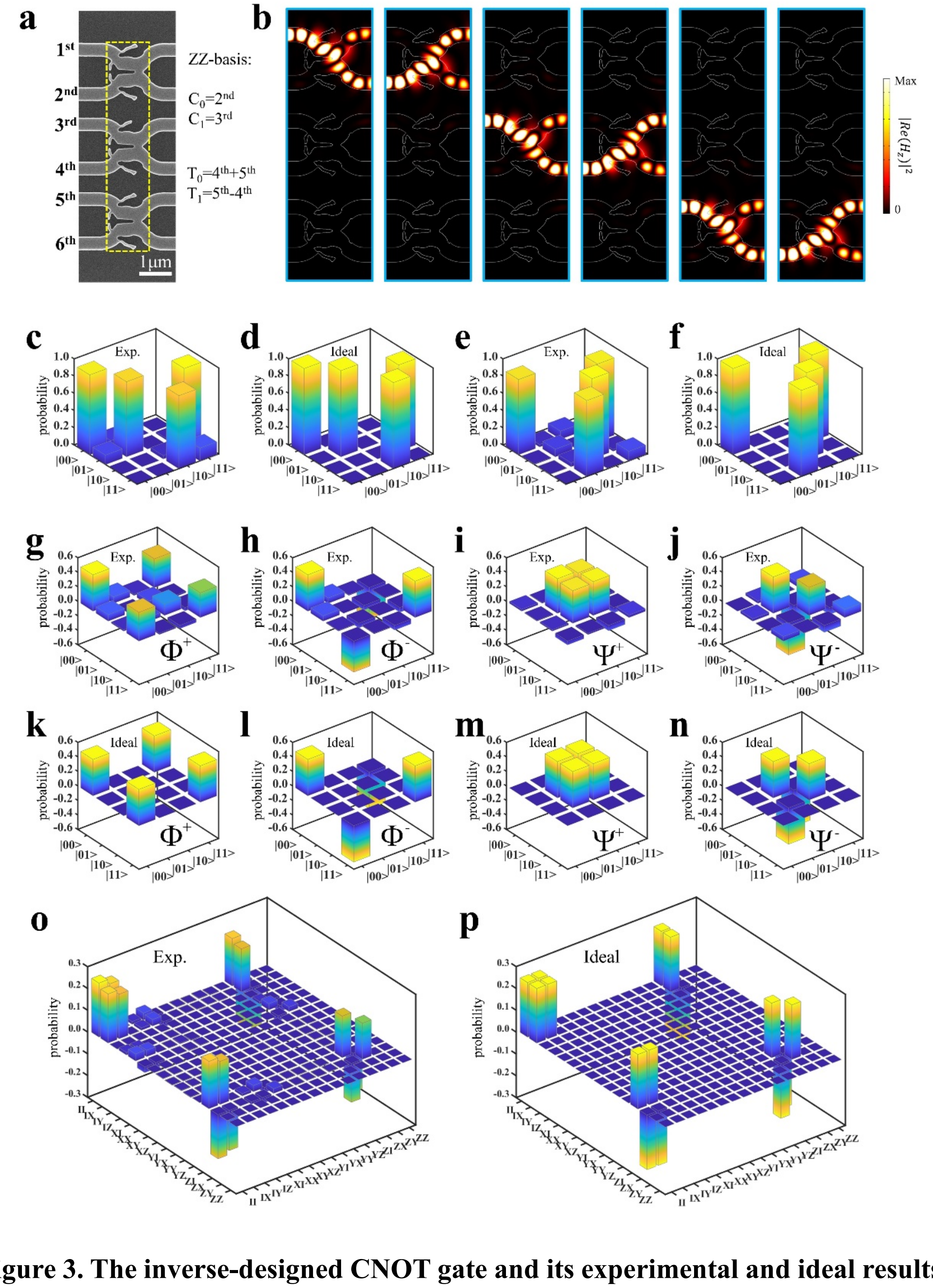

**Figure 3. The inverse-designed CNOT gate and its experimental and ideal results. a.** The SEM image of the CNOT gate. **b.** The simulation results of three 33:67 beam splitters in the CNOT gate. The **(c)** experimental and **(d)** ideal results of the operation probabilities for the CNOT gate under the ZZ-basis. The experimental fidelity is $F_{ZZ}$=0.9481±0.0064. The **(e)** experimental and **(f)** ideal results of the operation probabilities for the CNOT gate under the XX-basis. The experimental fidelity is $F_{XX}$=0.9445±0.0051. **g-j.** The experimental density matrices of four bell states, the fidelities $F_{Bell}$ are 0.9034±0.0110, 0.9634±0.0059, 0.9578±0.0068, and 0.9382±0.0067 respectively. **k-n.** The ideal density matrices of four bell states. The quantum process

tomography of the CNOT gate for **(o)** the experimental results and **(p)** the ideal results. The process fidelity $F_{CNOT}$ is 0.9080±0.0030.

In our inverse-designed CNOT gate, there are six input and output waveguides, named from 1st to 6th, connecting to three 33:67 BSs from top to bottom of Fig. 3a. For simplicity, the quantum states in these input waveguides are defined as $|1^{st}\rangle$, $|2^{nd}\rangle$, $|3^{rd}\rangle$, $|4^{th}\rangle$, $|5^{th}\rangle$, and $|6^{th}\rangle$. Fig. 3b shows the simulated field distributions when these single-photon states with λ=1550nm are injected into the waveguides, respectively. The devices possess the nice performances to implement the function of the 33:67 BSs with low loss, and no crosstalk coupling among each other.

To carry out the function of the CNOT gate, we consider the measurement of the operation under **ZZ**-basis and **XX**-basis. The **ZZ**-basis is defined as $|0_{ZZ}\rangle_c \equiv |2^{nd}\rangle$ and $|1_{ZZ}\rangle_c \equiv |3^{rd}\rangle$ for the control qubit, as well as $|0_{ZZ}\rangle_t \equiv \frac{1}{\sqrt{2}}(|4^{th}\rangle + |5^{th}\rangle)$ and $|1_{ZZ}\rangle_t \equiv \frac{1}{\sqrt{2}}(|5^{th}\rangle - |4^{th}\rangle)$ for the target qubit. To characterize the operation of this gate, we measure the output for each of the four possible input states: $|00_{ZZ}\rangle_{ct}$, $|01_{ZZ}\rangle_{ct}$, $|10_{ZZ}\rangle_{ct}$, and $|11_{ZZ}\rangle_{ct}$. The measured results for input-output operation probabilities, normalized by the sum of all coincidence counts obtained for each of the respective input states, are presented in Fig. 3c. The corresponding ideal result is shown in Fig. 3d. By comparison, we find that the experiment result is in good agreement with the theory, indicating a nice performance of the designed CNOT gate. Furthermore, the average transformation fidelity of the CNOT gate can be obtained as $F_{ZZ}$=0.9481±0.0064, where the definition of this fidelity is $F_{ZZ} = Tr(\sqrt{\sqrt{M_{th}} M_{exp} \sqrt{M_{th}}})$. Then, let us consider the XX-basis, which is defined as $|0_{XX}\rangle_c \equiv \frac{1}{\sqrt{2}}(|2^{nd}\rangle + |3^{rd}\rangle)$ and $|1_{XX}\rangle_c \equiv \frac{1}{\sqrt{2}}(|2^{nd}\rangle - |3^{rd}\rangle)$ for the control qubit, as well as $|0_{XX}\rangle_t \equiv |5^{th}\rangle$ and $|1_{XX}\rangle_t \equiv |4^{th}\rangle$ for the target qubit. As shown in Fig. 3e, the operation of the gate presents the correct output states $|00_{XX}\rangle_{ct}$, $|01_{XX}\rangle_{ct}$, $|10_{XX}\rangle_{ct}$, and $|11_{XX}\rangle_{ct}$ corresponding to the input states $|00_{XX}\rangle_{ct}$, $|10_{XX}\rangle_{ct}$, $|00_{XX}\rangle_{ct}$, and $|01_{XX}\rangle_{ct}$ respectively, which is also in good agreement with the theoretical data in Fig. 3f. The average transformation fidelity can be computed as $F_{xx}$=0.9445±0.0051. Such high fidelities under **ZZ**- and **XX**-basis quantitatively confirm that implements the quantum CNOT function well. The small discrepancy between the experimental and ideal fidelities is mainly attributed to the inaccuracy of the quantum state preparing and tomography on the chip, and the imperfect BS ratio for the 33:67 BSs. It should be noted that, due to the post-selection strategy, the success probability of the coincidence measurement for the CNOT gate is theoretically 1/9.

An important function of the CNOT gate is to entangle two quantum states. In a particular case, maximally entangled Bell states $\Phi^+$, $\Phi^-$, $\Psi^+$ and $\Psi^-$ can be generated by inputting the superposition states $|\pm\rangle_c|0\rangle_t$ and $|\pm\rangle_c|1\rangle_t$, where $|\pm\rangle_c=\frac{1}{\sqrt{2}}(|0\rangle_c\pm|1\rangle_c)$. In the experiment, the corresponding quantum states $\frac{1}{\sqrt{2}}(|2^{nd}\rangle\pm|3^{rd}\rangle)$ and $\frac{1}{\sqrt{2}}(|4^{th}\rangle\pm|5^{th}\rangle)$ are injected into the CNOT gate. And then, we employ the arbitrary single-qubit measurement of the capability to analyze these four output states for performing the quantum state tomography. The phase shifters are adjusted to implement all the measurements in the state preparing and tomography modules of the chip. Therefore, the corresponding density matrix of Bell states can be reconstructed by quantum state process tomography (*47*). Figs. 3g-3j show the experimentally measured density matrices of bell states. Correspondingly, the ideal density matrices of bell states are given in Figs. 3k-3n. It can be seen that all four Bell states are accurately generated. The corresponding fidelities of bell states are obtained from $F_{Bell}=\mathrm{Tr}(\sqrt{\sqrt{\rho_{th}}\rho_{exp}\sqrt{\rho_{th}}})$ as 0.9034±0.0110, 0.9634±0.0059, 0.9578±0.0068, and 0.9382±0.0067, respectively. Here, $\rho_{exp}$ ($\rho_{th}$) is the density matrix reconstructed from the experimental (ideal) data. The results of high fidelities show that four Bell states are well generated by the inverse-designed CNOT gate, which demonstrates the entanglement ability of the gate.

To fully characterize the inverse-designed CNOT gate, we also carry out the quantum process tomography. For a generic quantum process ζ acting on a 2-qubit density matrix β, one has $\zeta(\beta)=\sum_{m,n=1}^{15}\chi_{mn}\hat{A}_m\beta\hat{A}_n^\dagger$, where the operator $\hat{A}_m$ ($\hat{A}_n$) is defined as the tensor products of Pauli matrices {$\hat{A}_m\equiv\sigma_i\otimes\sigma_j$}, i, j=0,…,3, m=0,…,15. Here, the matrix $\chi_{mn}$ contains all the information of the process. The experimentally reconstructed process matrix is plotted in Fig. 3o, which is basically consistent with the ideal case as shown in Fig. 3p. Using the definition of the process fidelity $F_{exp} = \mathrm{Tr}\left[\sqrt{\sqrt{\chi_{exp}}\chi_{CNOT}\sqrt{\chi_{exp}}}\right]^2 \Big/ \mathrm{Tr}\left[\chi_{exp}\right]\mathrm{Tr}\left[\chi_{CNOT}\right]$, we obtain $F_{exp}$=0.9080±0.0030, which shows a high-performance efficiency of the designed CNOT gate. In addition, our experiment results also show very small imaginary parts (close to zero) for the density matrices of the bell states and the reconstructed process matrix of the CNOT gate. This further demonstrates the nice performance of our CNOT gate. The detailed discussion can be found in S5 of the Supplementary Materials.

## 3. Discussion

### 3.1 The discussion of super-compact quantum circuits.

One of the major applications of quantum gates is the construction of integrated quantum photonic circuits to implement arbitrary quantum processing. Thus, it is of great significance to demonstrate that our inverse-designed quantum gates can be used to construct such a quantum circuit with a super-compact footprint. As the schematic diagram shown in Fig. 4a, the arbitrary 2-qubit quantum circuit consists of three inverse-designed CNOT gates and eight arbitrary R gates. The footprint of such a quantum circuit is approximately $10^3\mu m^2$. It is effectively shrunk 4 orders of magnitude from ~$10^7\mu m^2$ in the previous work (*4*), which indicates that more than $10^4$ 2-qubit quantum circuits can be integrated into the same area. The overall size of the proposed quantum circuit can be found in S6 of Supplementary Materials.

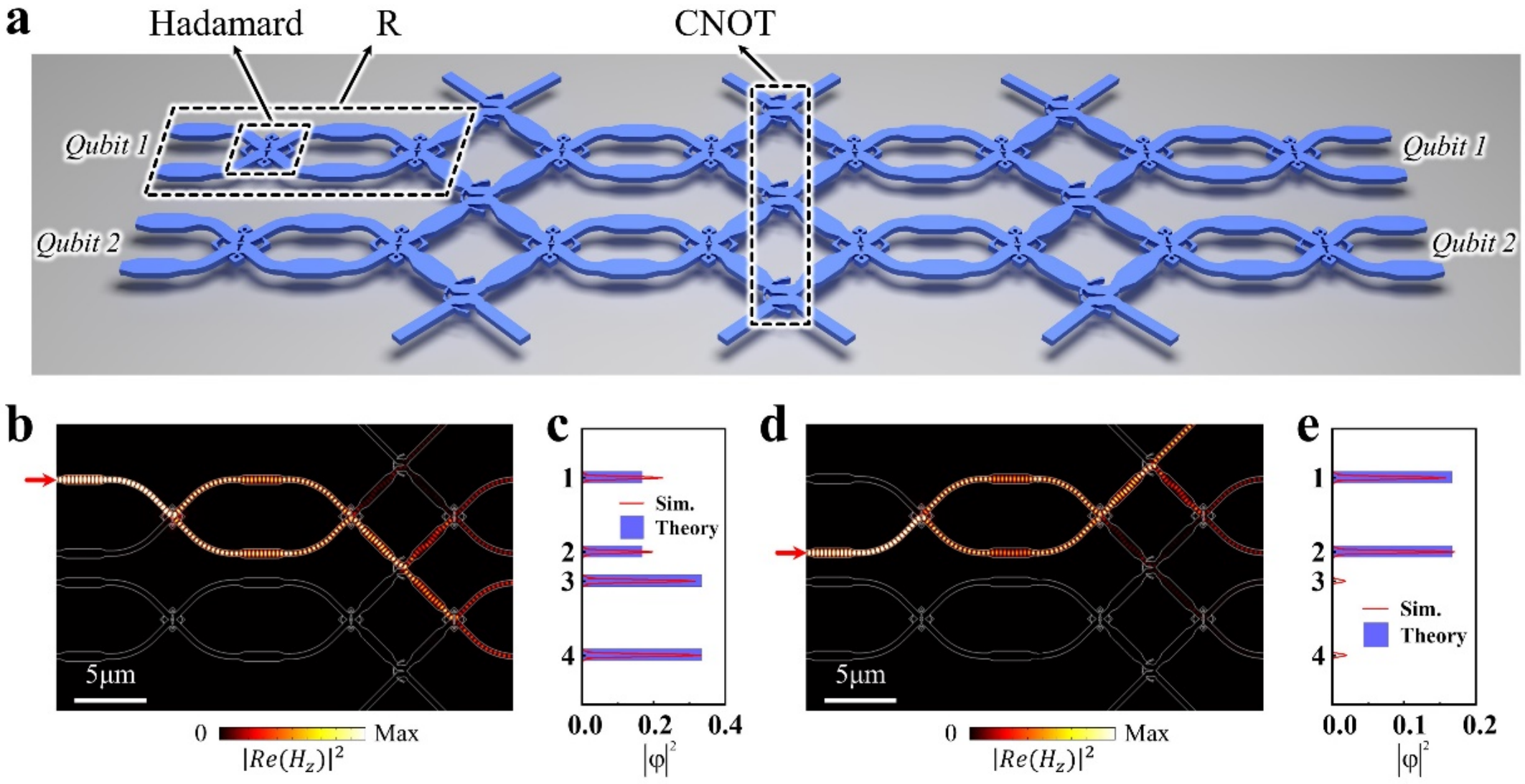

**Figure 4. The scheme of inverse-designed super-compact quantum circuits. a.** The schematic diagram for inverse-designed super-compact quantum circuits consisted of eight arbitrary single-qubit gates (R) and three CNOT gates. **b** and **d** represent the simulation results of the field distribution for the quantum circuit with single-photon exciting from the **(b)** first and **(d)** second waveguides. **c** and **e** are the comparison figures between theory (blue rectangles) and simulation (red lines) results of the single-photon state probability, where the theory results are calculated by the transfer matrix method and the simulation results are obtained by the field distribution in the waveguides.

In order to test the function of our designed super-compact quantum circuits, we perform a numerical simulation of single-photon state evolution. Figs. 4b and 4d show squares of probability amplitudes ($|\varphi|^2$) with the single-photon state exciting from the first and second waveguides, respectively. The corresponding $|\varphi|^2$ of output states from the first to the fourth waveguides is marked as the red lines in Figs. 4c and 4e. Meanwhile, we also provide theoretical results for the squares of probability amplitudes of the output superposition state in these four waveguides, which are marked as blue rectangles. The detailed theoretical method is described in S7 of the Supplementary Materials. Comparing them, we find that the consistency between the theory and the numerical simulation is very well, which indicates that the circuit has a high performance and low crosstalk even though it is integrated into such a small footprint. Furthermore, the state transform matrices of the quantum circuit are also calculated, the consistency between the theoretical and numerical results is proved again. This means that quantum chips with good functions can indeed be fabricated by the inverse design method. In fact, for some specific 2-qubit processes, such as the SWAP gate, a more compact footprint (~50μm$^2$) can be achieved (see S8 of Supplementary Materials for details).

After 2-qubit universal quantum logic gates have been designed, the next milestone goal is to realize the ultra-small quantum computing chip for some certain tasks, such as boson sampling (*48, 49*). For these tasks, variable optical elements are more meaningful, which is primarily a thermally-controlled phase shifter. We believe it can also be optimized by the inverse design method to obtain ultra-compact footprints. Furthermore, it can greatly reduce the size of tunable quantum chips. In general, the overall fidelity of a quantum circuit decreases with the number of device cascades increasing. Thus, the high fidelity of the unit device is particularly important for large-scale cascaded quantum circuits. Recent works (*50-52*) report that single device fidelity of more than 99% enables fault-tolerant quantum computing. In the future. we will take the fidelity as the objective function in the optimization, to obtain high fidelity quantum gates for large-scale quantum circuits.

Overall, it needs to solve the problems of the full tunability of an optical chip in ultra-small footprints, cross-talks of quantum states in such a small scale, and fabricated technique of silicon photonic chip with low-loss inverse-designed devices. After these problems are solved, we believe that the ultra-small quantum computing chip can be realized.

In summary, we have designed and fabricated super-compact universal quantum logic gates using the inverse-designed method on the silicon photonic chip with the integrated source. The footprints of CNOT and Hadamard gate are only 8.32μm$^2$ (1.3μm×6.4μm) and 1.69μm$^2$ (1.3μm×1.3μm), respectively. They are the smallest optical quantum gates reported until now. Based on these universal quantum logic gates, the silicon photonic quantum circuit to implement arbitrary two-qubit processing has been designed. It is found that the size of the quantum circuit is reduced by 4 orders of magnitude compared with those previous quantum photonic circuits. The high-performance efficiencies for these super-compact quantum gates and circuits have also been demonstrated. This work provides some novel designs for on-chip integrated quantum information processing, which is expected to solve the scalability problem of optical quantum chips.

## 4. Methods

### 4.1 Measurement method.

The continuous wave laser (keysight N7714A) is employed to generate the pump light in the experiment, and the wavelengths of the two pump lights are 1538.19nm and 1535.04nm, respectively. The incident laser is first coupled to the single mode fibers (SMF), combined by the DWDMs, and injected into the chip by the fiber array. Next, the SFWM process is stimulated in two 6mm-length silicon waveguides, where the frequency-degenerate photon pairs (1536.61nm) are generated. The two-photon coincidence counts are up to 100kHz under the pump power of 35mW. Nine computer-controlled thermal phase shifters are used for generating any biphotonic state and projecting it to any measurement basis. The output photons from the chip coupled to the SMFs of the fiber array, went through the DWDMs (filter out the pump light), and are counted by the SNSPDs. Finally, we perform a coincidence measurement between the two photons.

The success probability of the quantum logic gate is determined by the collection efficiency of photons, which is constructed by the detection efficiency of photons and the coupling efficiency of the grating coupler. In the experiment, we exactly adjusted the position of the fiber array. The coupling efficiency of the grating coupler can reach -3.5dB for every coupler at the optimum wavelength. The efficiencies of single photon detectors are about ~50% and dark count rates are about ~100 Hz. In the future, we will further increase the collection efficiency to obtain a higher success probability. Potential methods include the utilization of the edge coupling and the SNSPDs with a better performance.

**4.2 Sample fabrication.**

The samples are fabricated using electron beam lithography, followed by dry etching. The substrate is a silicon-on-insulator wafer with a 220 nm-thick top Si layer. ZEP-520A e-beam resist is first spin-coated on the substrate for exposure, and resist patterns are formed after e-beam lithography and development. The fabrication time is about tens of minutes. Then these resist patterns are transformed to the top Si layer using inductively coupled plasma etching in SF6 and CHF3 gases atmosphere, with ZEP520A used as an etching mask. The etching depth for inverse-designed structures and waveguides is 220nm. Next, a 1μm-thick silicon dioxide ($SiO_2$) layer is deposited by plasma-enhanced chemical vapor deposition (PECVD). Finally, a layer of 100 nm-thick titanium (Ti) is deposited on top of waveguides to form thermal-optical phase-shifters. Moreover, the photolithography tool can also be used to fabricate our device, which can greatly reduce the fabrication time. It is more practical for future large-scale integration. Photon pairs are generated in silicon waveguides with a 450nm width, 220nm height, and 6mm length, by the SFWM nonlinear process. MMIs with a width of 6μm and length of 43μm are used as balanced beamsplitters with a low loss (less than 0.5dB).

**Supplementary Materials**

Section S1. The size of each area in the quantum chip for measuring the CNOT gate.
Section S2. The detailed optimization procedures of Hadamard and CNOT gates.
Section S3. The detailed discussion of the sensitivity analysis of the device performance.
Section S4. The detailed discussion of the arbitrary single qubit gate.
Section S5. The imaginary parts for the density matrices of the bell states and the reconstructed process matrix of the CNOT gate.
Section S6. The overall size of the proposed quantum circuit.
Section S7. The simulation details of the optical transform matrix.
Section S8. The inverse-designed SWAP gate.

**References and Notes**

1. A. Barenco, et al. Elementary gates for quantum computation *Phys. Rev. A*, **52**, 3457–3467 (1995).
2. S. Daiss et al. A quantum-logic gate between distant quantum-network modules *Science*, **371**, 614-617 (2021).
3. D. Hanneke, J. P. Home, J. D. Jost, J. M. Amini, D. Leibfried and D. J. Wineland, Realization of a programmable two-qubit quantum processor *Nat. Physics*, **6**, 13 (2010).

4. J. Wang et al. Integrated photonic quantum technologies *Nat. Photonics*, **14**, 273-284 (2020).
5. X. Qiang et al. Large-scale silicon quantum photonics implementing arbitrary two-qubit processing *Nature photonics*, **12**, 534-539 (2018).
6. J. L. O'Brien, A. Furusawa, J. Vučković, Photonic quantum technologies *Nat. Photonics* **3**, 687 (2009).
7. A. Politi, M. J. Cryan, J. G. Rarity, S. Yu, J. L. O'brien, Silica-on-silicon waveguide quantum circuits *Science* 320, 646 (2008).
8. X. Li et al. An all-optical quantum gate in a semiconductor quantum dot. *Science*, **301**, 809-811 (2003).
9. C. D. Hill et al. A surface code quantum computer in silicon. *Science advances*, 1, e1500707 (2015).
10. R. Okamoto et al. Demonstration of an Optical Quantum Controlled-NOT Gate without Path Interference *Phys. Rev. Lett.* **95**, 210506 (2005).
11. T. C. Ralph et al. Simple scheme for efficient linear optics quantum gates *Phys. Rev. A* **65**, 012314. (2001).
12. N. J. Cerf, Adami, P. G. Kwiat, Optical simulation of quantum logic *Phys. Rev. A* **7**, 1477 (1998).
13. H. F. Hofmann, S. Takeuchi, Quantum phase gate for photonic qubits using only beam splitters and postselection *Phys. Rev. A* **66**, 024308 (2002).
14. A. S. Clark et al. All-optical-fiber polarization-based quantum logic gate *Phys. Rev. A* **79**, 030303. (2009).
15. E. Knill, R. Laflamme, G. J. Milburn, A scheme for efficient quantum computation with linear optics *Nature* **409**, 46-52 (2001).
16. F. Schmidt-Kaler et al. Realization of the Cirac-Zoller controlled-NOT quantum gate *Nature* **422**, 408-411 (2003).
17. A. Crespi et al. Integrated photonic quantum gates for polarization qubits *Nat. Commun.* **2**, 1-6 (2011).
18. J. Zeuner et al. Integrated-optics heralded controlled-NOT gate for polarization-encoded qubits *npj Quantum Inf.* **4** 1-7 (2018).
19. S. M. Wang et al. A 14×14 $\mu m^2$ footprint polarization-encoded quantum controlled-NOT gate based on hybrid waveguide. *Nat. Commun.* **7**, 1-5 (2016).
20. M. Zhang et al. Supercompact Photonic Quantum Logic Gate on a Silicon Chip *Phys. Rev. Lett.* **126**, 130501 (2021).
21. R. B. Patel, J. Ho, F. Ferreyrol, T. C. Ralph, G. J. Pryde, A quantum Fredkin gate. *Science advances*, **2**, e1501531 (2016).
22. D. Tiarks, S. Schmidt, G. Rempe, S. Dürr, Optical π phase shift created with a single-photon pulse. *Science Advances*, **2**, e1600036 (2016).
23. Y. Zhang, F. S. Roux, T. Konrad, M. Agnew, J. Leach, and A. Forbes, Engineering

two-photon high-dimensional states through quantum interference. *Science advances*, **2**, e1501165 (2016).

24. R. Uppu et al. Scalable integrated single-photon source. *Science advances*, **6**, eabc8268 (2020).
25. S. Molesky et al. Inverse design in nanophotonics *Nat. Photonics* **12**, 659-670 (2018).
26. J. S. Jensen, O. Sigmund, Topology optimization for nano-photonics *Laser Photonics Rev.* **5**, 308-321 (2011).
27. A. Y. Piggott et al. Inverse design and demonstration of a compact and broadband on-chip wavelength demultiplexer *Nat. Photonics* **9**, 374-377 (2015).
28. B. Shen et al. An integrated-nanophotonics polarization beamsplitter with 2.4×2.4 $\mu m^2$ footprint *Nat. Photonics* **9**, 378-382 (2015).
29. N. Mohammadi Estakhri, B. Edwards, N. Engheta, Inverse-designed metastructures that solve equations *Science* **363**, 1333-1338 (2019).
30. M. Camacho, B. Edwards, N. Engheta, A single inverse-designed photonic structure that performs parallel computing *Nat. Commun.* **12**, 1-7 (2021).
31. N. V. Sapra et al. On-chip integrated laser-driven particle accelerator *Science* **367**, 79-83. (2020).
32. O. Yesilyurt et al. Efficient Topology Optimized Couplers for On-Chip Single-photon Sources *ACS Photonics* **8**, 3061-3068 (2021).
33. K. Yao, D. R. Unni, Y. Zheng, Intelligent nanophotonics: merging photonics and artificial intelligence at the nanoscale *Nanophotonics* **8**, 339–366 (2019).
34. Su, L. et al. Nanophotonic Inverse Design with SPINS: Software Architecture and Practical Considerations *Appl. Phys. Rev.* **7**, 011407 (2020).
35. L. Su, A. Y. Piggott, N. V. Sapra, J. Petykiewicz, J. Vučković, Inverse Design and Demonstration of a Compact on-Chip Narrowband Three-Channel Wavelength Demultiplexer *ACS Photonics* **5**, 301–305 (2017).
36. C. Sitawarin, W. Jin, Z. Lin, A. W. Rodriguez, Inverse-designed photonic fibers and metasurfaces for nonlinear frequency conversion *Photonics Res.* **6**, B82 (2018).
37. T. W. Hughes, M. Minkov, I. A. D. Williamson, S. Fan, Adjoint Method and Inverse Design for Nonlinear Nanophotonic Devices *ACS Photonics* **5**, 4781–4787 (2018).
38. Z. Liu et.al. Integrated nanophotonic wavelength router based on intelligent algorithm *Optica* **6**, 1367-1373 (2019).
39. C. Dory et al. Inverse-designed diamond photonics *Nat. Commun.* **10**, 3309 (2019).
40. K. Y. Yang et al. Inverse-designed non-reciprocal pulse router for chip-based LiDAR *Nat. Photonics* **14**, 369–374 (2020).
41. John Peurifoy et al. Nanophotonic particle simulation and inverse design using artificial neural networks *Science advances* **4**, eaar4206 (2018).
42. Haoran Ren et al. Three-dimensional vectorial holography based on machine

learning inverse design *Science advances* **6**, eaaz4261 (2020).
43. S. H. Nam et al. Photolithographic realization of target nanostructures in 3D space by inverse design of phase modulation *Science advances*, **8**, eabm6310 (2022).
44. R. E. Christiansen and Ole Sigmund, Inverse design in photonics by topology optimization: tutorial *J. Opt. Soc. Am. B* **38**, 496-509 (2021).
45. R. E. Christiansen, B. S. Lazarov, J. S. Jensen, O. Sigmund, Creating geometrically robust designs for highly sensitive problems using topology optimization - acoustic cavity design *Structural and Multidisciplinary Optimization* **52**, 737 – 754 (2015).
46. R. Heilmann et al. Arbitrary photonic wave plate operations on chip: Realizing Hadamard, Pauli-X, and rotation gates for polarisation qubits *Sci. Rep.* **4**, 1-5 (2014).
47. L. Sansoni et al. Polarization Entangled State Measurement on a Chip *Phys. Rev. Lett.* **105**, 200503 (2010).
48. H.-S. Zhong et al. Quantum computational advantage using photons *Science* **370**, 1460–1463 (2020).
49. A. Deshpande et al. Quantum computational advantage via high-dimensional Gaussian boson sampling *Science advances*, **8**, eabi7894 (2022).
50. M. T. Mądzik et al. Precision tomography of a three-qubit donor quantum processor in silicon *Nature*, **601**, 348-353 (2022).
51. Xue, X. et al. Quantum logic with spin qubits crossing the surface code threshold *Nature*, **601**, 343-347 (2022).
52. A.Noiri, et al. Fast universal quantum gate above the fault-tolerance threshold in silicon *Nature*, 601, 338-342 (2022).

**Acknowledgements:** We would like to thank Hao Li and Lixing You from Shanghai Institute of Microsystem and Information Technology, Chinese Academy of Sciences, China for their support on the SNSPDs used in this work. L. He thanks Rasmus E. Christiansen for helpful discussions. This work was supported by the National key R & D Program of China (2017YFA0303800, 2018YFB2200400), National Natural Science Foundation of China (No.91850205) and Beijing Natural Science Foundation (Z180012).

**Author contributions:** L. He and D. Liu contributed equally to this work. L. He finished the theoretical scheme and designed the experiments. L. He finished experiments with the help of D. Liu, J. Gao, H. Zhang, X. Feng, F. Liu, K. Cui under the supervision of W. Zhang and Y. D. Huang. L. He and X. D. Zhang wrote the paper with the help of W. Zhang, D. Liu and W. X. Zhang. X. D. Zhang initiated and designed this research project.

**Competing interests:** The authors declare no competing interests.

**Data availability:** All data needed to evaluate the conclusions in the paper are present in the paper and/or the supplementary materials

# Supplementary Materials for

## Super-compact universal quantum logic gates with inverse-designed elements

Lu He[1*], Dongning Liu[2*], Jingxing Gao[2], Weixuan Zhang[1], Huizhen Zhang[1], Xue Feng[2], Yidong Huang[2,3], Kaiyu Cui[2], Fang Liu[2], Wei Zhang[2,3+] and Xiangdong Zhang[1$]

[1]*Key Laboratory of advanced optoelectronic quantum architecture and measurements of Ministry of Education, Beijing Key Laboratory of Nanophotonics & Ultrafine Optoelectronic Systems, School of Physics, Beijing Institute of Technology, 100081 Beijing, China.*

[2]*Frontier Science Center for Quantum Information, Beijing National Research Center for Information Science and Technology (BNRist), Electronic Engineering Department, Tsinghua University, Beijing 100084, China*

[3]*Beijing Academy of Quantum Information Sciences, Beijing 100193, China.*

**These authors contributed equally to this work. [$+]Author to whom any correspondence should be addressed: zhangxd@bit.edu.cn; zwei@tsinghua.edu.cn*

**S1. The size of each area in the quantum chip for measuring the CNOT gate.**

Here, we take the CNOT gate as an example to show the chip design layout. As shown in Fig. S1a, the whole footprint of the chip is about 4.3mm×1.4mm. It can be seen that most of the chip is the supporting modules to test the super-compact quantum gate.

They correspond to the modules of (I) quantum source, (II) state preparing, (III) quantum gate, and (IV) state tomography, which are mentioned in Fig. 1a of the main text. The size of the quantum source part is about 1400μm×230μm. The size of the state preparing (tomography) part is about 900μm×400μm. The size of the quantum gate part is 1.3μm×6.4μm.

Our proposed super-compact quantum gate is marked in the blue box. The corresponding SEM image and enlarged SEM image can be seen in Figs. S1b and S1c. We have demonstrated that our quantum gate has a good performance through traditional devices (MZI, quantum source, etc.).

Through Fig. S1a, we can clearly compare the size of the inverse-designed quantum gate with that of traditional quantum devices. The size of traditional quantum devices is much larger than that of inverse-designed quantum devices. We expect to apply inverse-designed devices on the entire chip in the future, for reducing the overall size of quantum chips.

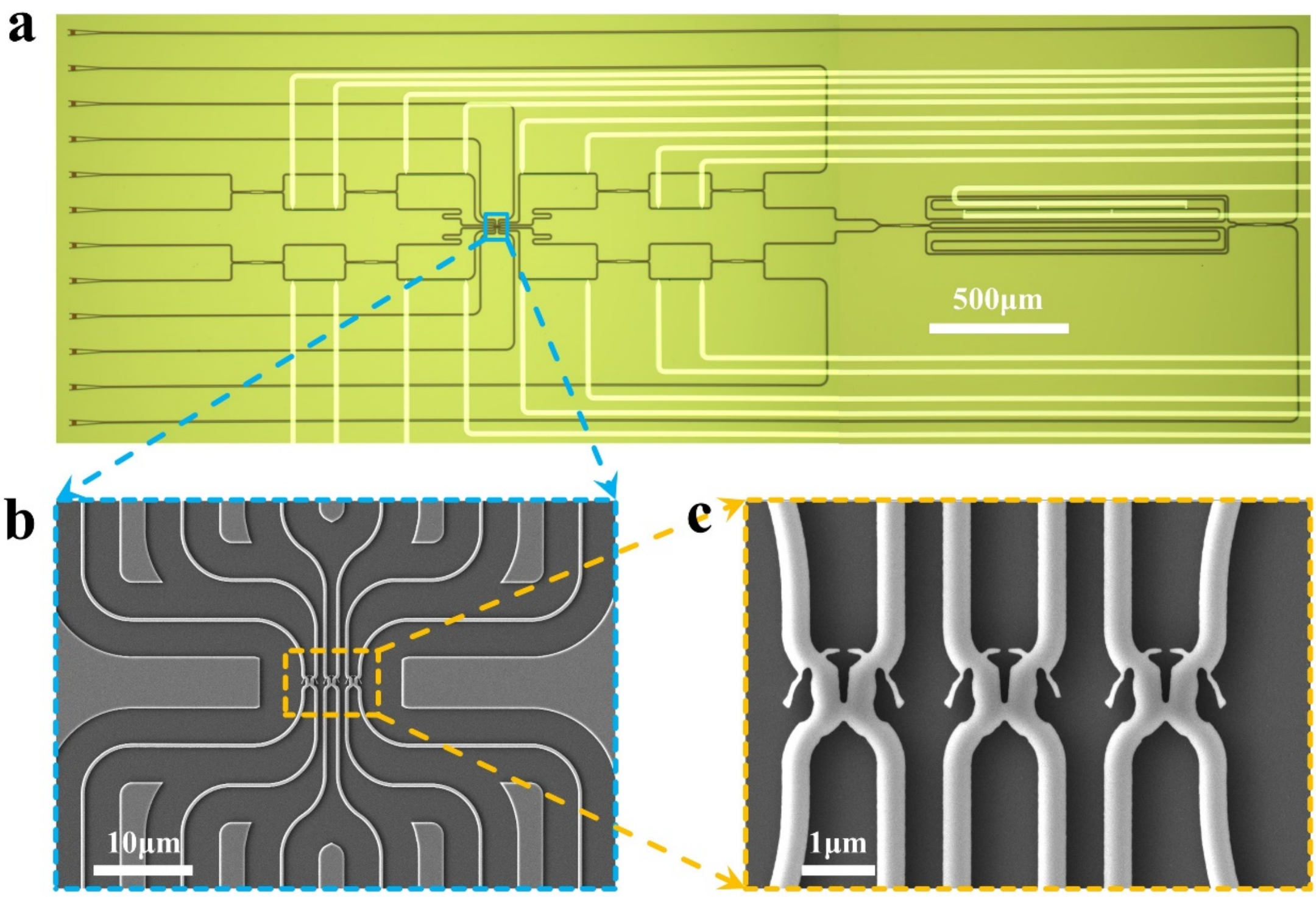

**Fig. S1. The quantum chip for measuring the CNOT gate. a**. The microscope picture of the quantum chip, which corresponds to Fig. 1a in the main text. Because the microscope field of view is relatively small, the whole microscope picture is taken by twice independent photographing. And then we joint them together. **b.** The SEM image of the CNOT gate and its connected waveguide. **c.** The enlarged SEM image of the CNOT gate.

## S2. The detailed optimization procedures of Hadamard and CNOT gates.

From the main text, it is known that the kernels of the Hadamard and CNOT gates are mainly formed by the 50:50 and 33:67 beam splitters for the linear optical scheme. Thus, in this section, we focus on the optimization of the 50:50 and 33:67 beam splitter in a single structure, where a pair of input ports and two output ports (named $a_{in}$, $b_{in}$ and $a_{out}$, $b_{out}$ ports) exist, as shown in Fig. S2a. Note that in our optimization process, the squares of the amplitude of the single photon quantum state corresponds to the optical field energy.

For the 50:50 beam splitter, if only one single photon is injected into the system from any input port ($a_{in}$ or $b_{in}$), the output photonic quantum state could become the superposition at two output ports ($a_{out}$ and $b_{out}$) with equal squares of amplitudes at the same time. To fulfill these operations, the system must possess a mirror symmetry with respect the dashed line (called mirror symmetry line), as marked in Fig. S2a. In this case, only a half region is needed to be optimized where the other mirror symmetric part should have the same dielectric distribution.

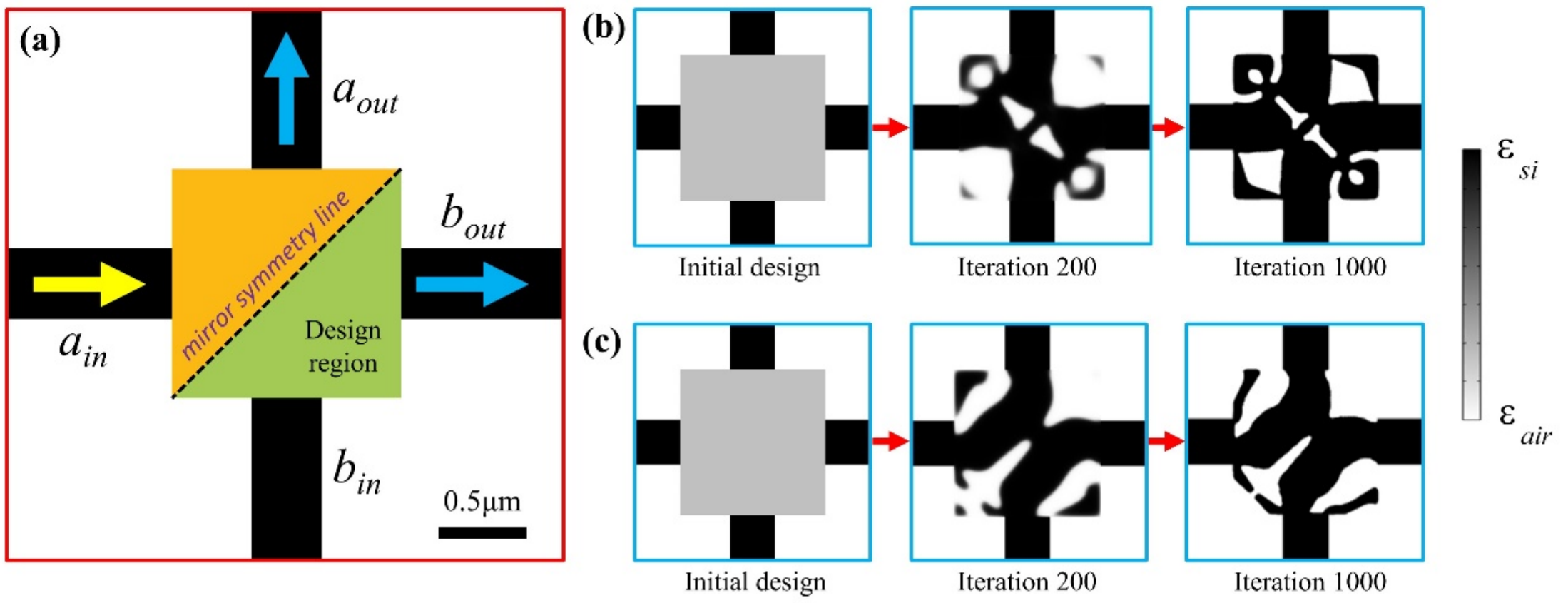

**Fig. S2. The inverse design process of the 50:50 and 33:67 beam splitters.** (a) the schematic diagram of the optimization model. The designable region (green triangle) and the mirror symmetry line (dashed line). The inverse design processes of (b) 50:50 and (c) 33:67 beam splitters. The initial design is an intermediate material between Si and $SiO_2$. With the optimization going on, the materials in the design region tend to binarization. The final design is determined by 1000 iterations.

To achieve the above-mentioned device, the associated objective function expressed as Eq. (1) in the main text is carried out. The detailed material parameters (the permittivities of $SiO_2$ and Si layer) are set as $\varepsilon_{sio_2}$=2.1 and $\varepsilon_{si}$=12.1. For the additional condition $\gamma_1 \le c_1^2/c_2^2 \le \gamma_2$, we set $\gamma_1$ =0.9 and $\gamma_2$=1.1 in the optimization process. In this way, the inverse design process is carried out from the initial design of the homogeneous material distribution with the dielectric constant being $\varepsilon=(\varepsilon_{sio_2}+\varepsilon_{si})/2$. The intermediate structures generated by the inverse design process are shown in Fig. S2b. With the increasing of iteration steps, the objective function is gradually enlarged and the final structure with the optimal performance appears after 1000 iterations.

For the 33:67 beam splitter, the inverse design process is similar to that of the 50:50 beam splitter, except for adjusting the parameters $\gamma_1$ and $\gamma_2$ in the extra condition. They are changed as $\gamma_1$ =0.62 and $\gamma_2$=0.72. Similarly, the intermediate structures generated by the inverse design process are shown in Fig. S2c. The corresponding final structure appears after 1000 iterations. Except for these limiting conditions, the linear material interpolation, projection, and filtering procedures are used in the inverse design process.

Finally, the structures of the optimized 50:50 and 33:67 beam splitters are exported for realizing the Hadamard and CNOT gates, respectively.

**S3. The detailed discussion of the sensitivity analysis of the device performance.**

In optical chip fabrication, the most common defect is the manufacturing deviation. Generally, there is an error of 20nm in the e-beam lithography etching, due to the instability of the dose of the electron beam. Thus, the designed device can be used for experimental fabrication if it can tolerate a manufacturing defect of under (over) etching with ±10nm.

Here, we consider the sensitivity analysis of the device performance with some manufacturing defects. Such an analysis includes two aspects: the first one is that we used the double filtering method (Refs. [45, 46] in the main text) in the inverse-design process, which can increase the robustness of the device; the second one is that we also did a simulation to verify the performance of our device. Let us introduce them in detail:

1) In order to increase the robust property against manufacturing defects, we use the double filtering method in the inverse-design process. In brief, the double filtering method consists of applying the filter and threshold procedure twice on the design field, where in the second application three different threshold values are applied to obtain three different realizations of the design fields corresponding to under (over) etching. So, the over and under etching cases are simultaneously optimized. In such a way, we can get a robust device against the manufacturing defect of under (over) etching, which is the most common defect in optical chip fabrication. Thus, the inverse-designed structure possesses the robust property against manufacturing defects in a certain extent.

2) After getting the inverse-designed structure from the optimization process, we further verify the device performance against manufacturing defects by the simulation. Here, we get three structures from the optimization results, which are exact etching (0nm) case and over or under etching (±10nm) cases. As shown in Fig. S3a, we selected 33:67 beam splitter as an example to perform sensitivity analysis. In the enlarged view of Fig. S3b, we can clearly see that the over or under etching (±10nm) is considered in the design layout. And then, we do the simulation (Figs. S3c-S3e) with the error of ±10nm. In these errors, the main function of the device (beam splitting ratio) deteriorate from 34:66 to 30:70 at the wavelength of 1550nm. The ideal beam splitting ratio is 33:67. The fidelity of the CNOT gate deteriorate from 99.96% to 99.36% in the theoretical calculation. A similar analysis about the required range of the splitting ratio can be found in Ref. [7] of the main text.

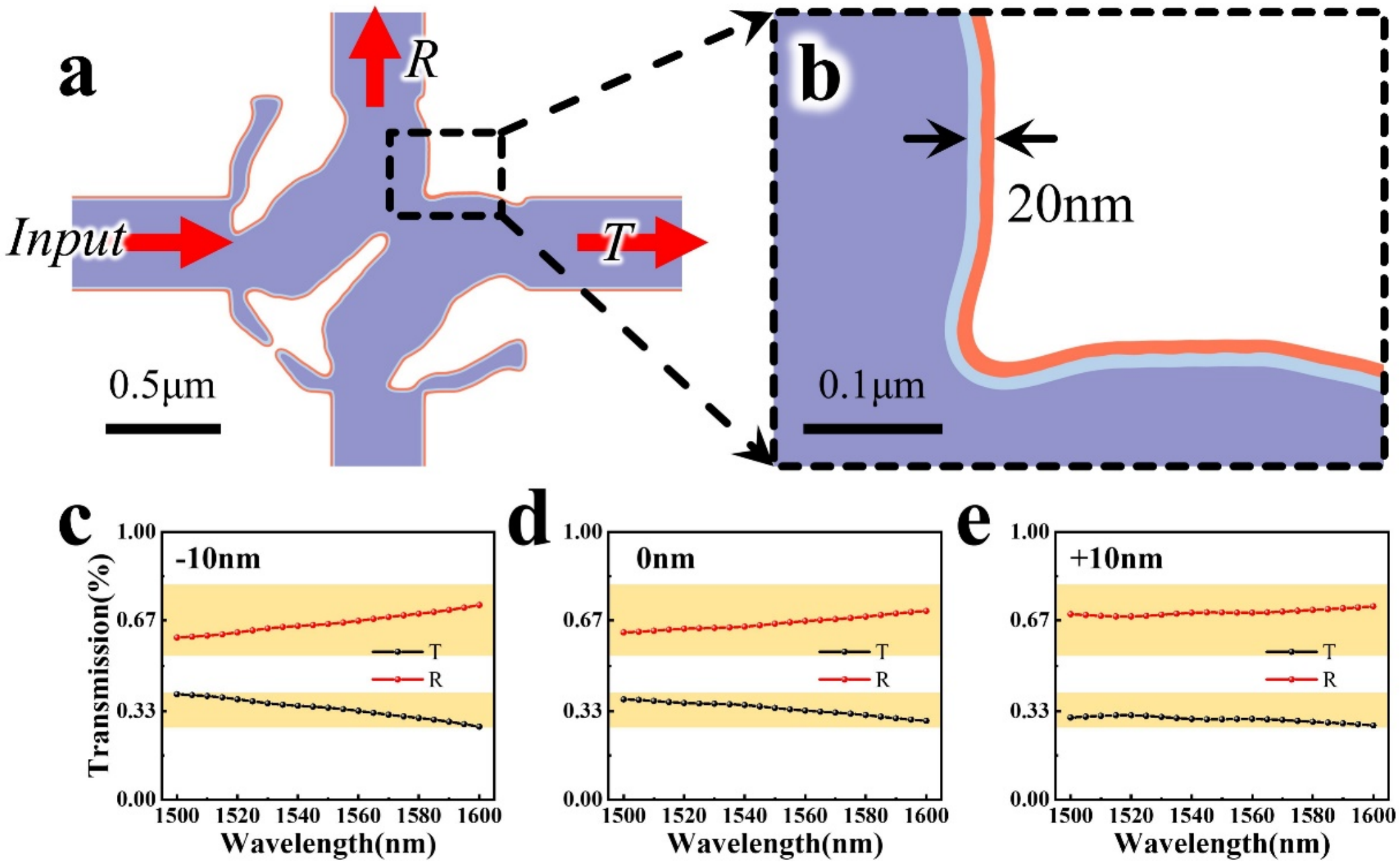

**Fig. S3. The sensitivity analysis of our designed device.** (a) The design layout with over (under) etching (±10nm). (b) The enlarged view of (a). The simulated results of the 33:67 beam splitter with (c) over etching (-10nm), exact etching, and under etching (+10nm). The yellow region represents the usable range of splitting ratio for the CNOT gate. For T, the range is from 0.60 to 0.73. For R, the range is from 0.27 to 0.40.

## S4. The detailed discussion of the arbitrary single qubit gate.

The single qubit gate consists of the Hadamard gate and the phase z gate. To ensure that their quantum circuits cascaded by individual gates work well, we must consider that the operating frequency range of these designed devices can cover each other.

First, we study the operating frequency range of the phase z gate by the simulation. We fix $L_2$=0.1μm and calculate the phase difference between the two waveguides after the gate by sweeping the parameter $L_1$, as shown in Fig. S4. It can be seen that in the whole communication band (1500-1600nm), the phase z gate can generate a stable phase difference θ by adjusting $L_1$.

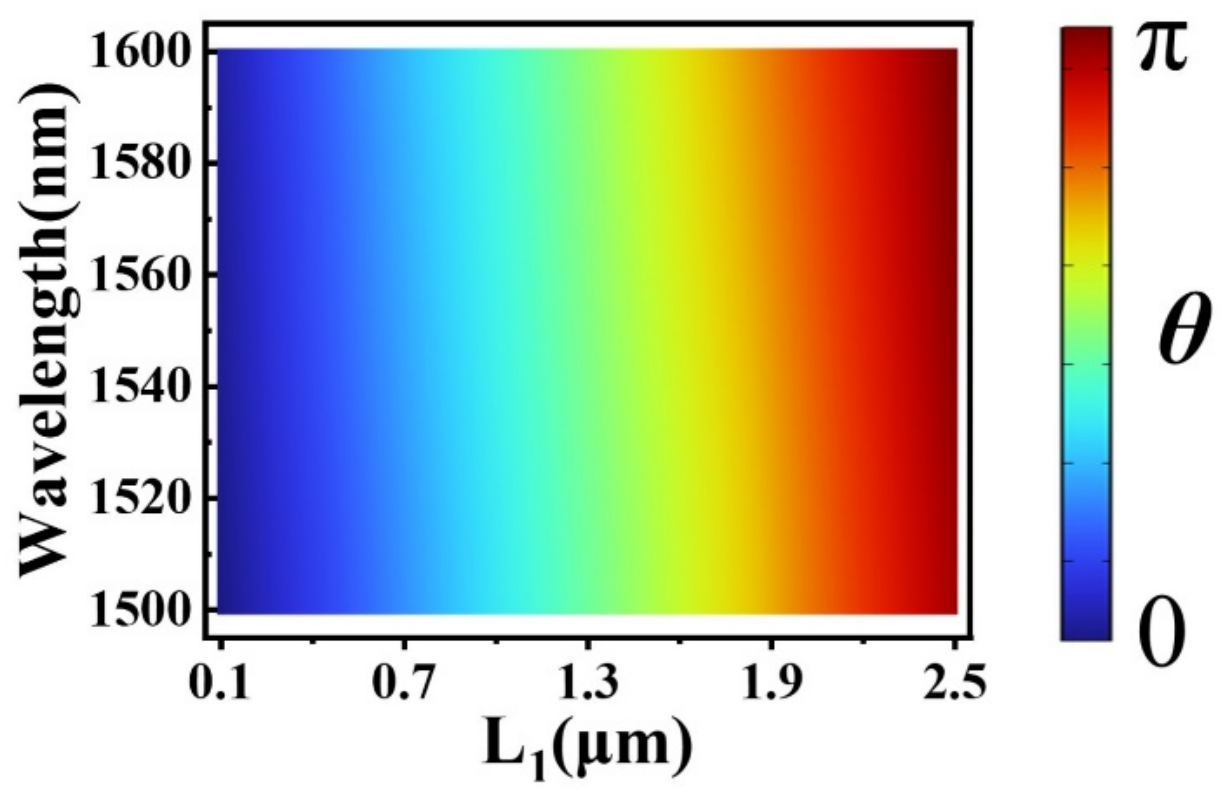

**Fig. S4. The additional phase difference of the phase z gate by adjusting $L_1$.**

And then, we study the theoretical model and the numerical simulation of the arbitrary single qubit gate R, as mentioned in Fig. 2g of the main text. For the path-encoded scheme, the R gate can be represented the product as a series of matrices. Thus, the quantum state evolution can be expressed as follow:

$$|\varphi_{out}\rangle = \frac{1}{2}\begin{pmatrix} 1 & \\ & e^{i\theta_3} \end{pmatrix}\begin{pmatrix} 1 & 1 \\ 1 & -1 \end{pmatrix}\begin{pmatrix} 1 & \\ & e^{i\theta_2} \end{pmatrix}\begin{pmatrix} 1 & 1 \\ 1 & -1 \end{pmatrix}\begin{pmatrix} 1 & \\ & e^{i\theta_1} \end{pmatrix}|\varphi_{in}\rangle. \tag{S2}$$

When a quantum state $|\varphi_{in}\rangle$ is injected into the single qubit gate R, any unitary transformation can be acted on it by adjusting the phase of $\theta_1$, $\theta_2$, and $\theta_3$. Now, we give three examples to demonstrate its function through the numerical simulations. In these three cases, the fixed phases in R gates are determined as $\theta_1=\pi$, $\theta_2=\pi/2$, and $\theta_3=0$, respectively. Hence, based on Eq. (S1), the output qubits a, b, and c become $|0\rangle$, $\frac{1}{\sqrt{2}}(|0\rangle + i|1\rangle)$, and $|1\rangle$, where $|0\rangle$ and $|1\rangle$ are the quantum state in the different paths. As shown in Figs. S5a-S5c, these three cases are simulated with the single photon exciting from the upper waveguide, where the quantum state is defined as $|0\rangle$. Correspondingly, the quantum state in the lower waveguide is $|1\rangle$. We find that the output quantum states indeed become the expected states after going through the R gate. In Fig. S5d, we mark these output quantum states to three points on the Bloch sphere. By the way, an arbitrary output quantum state can also be generated by the R gate, corresponding to the arbitrary point on the Bloch sphere. So, the simulation results possess a nice consistency with the theory results. It indicates that the R gate has a high performance even though that is integrated into such a small footprint. The overall size of the single-qubit quantum circuit in Fig. S5a-S5c is about 20μm×10μm.

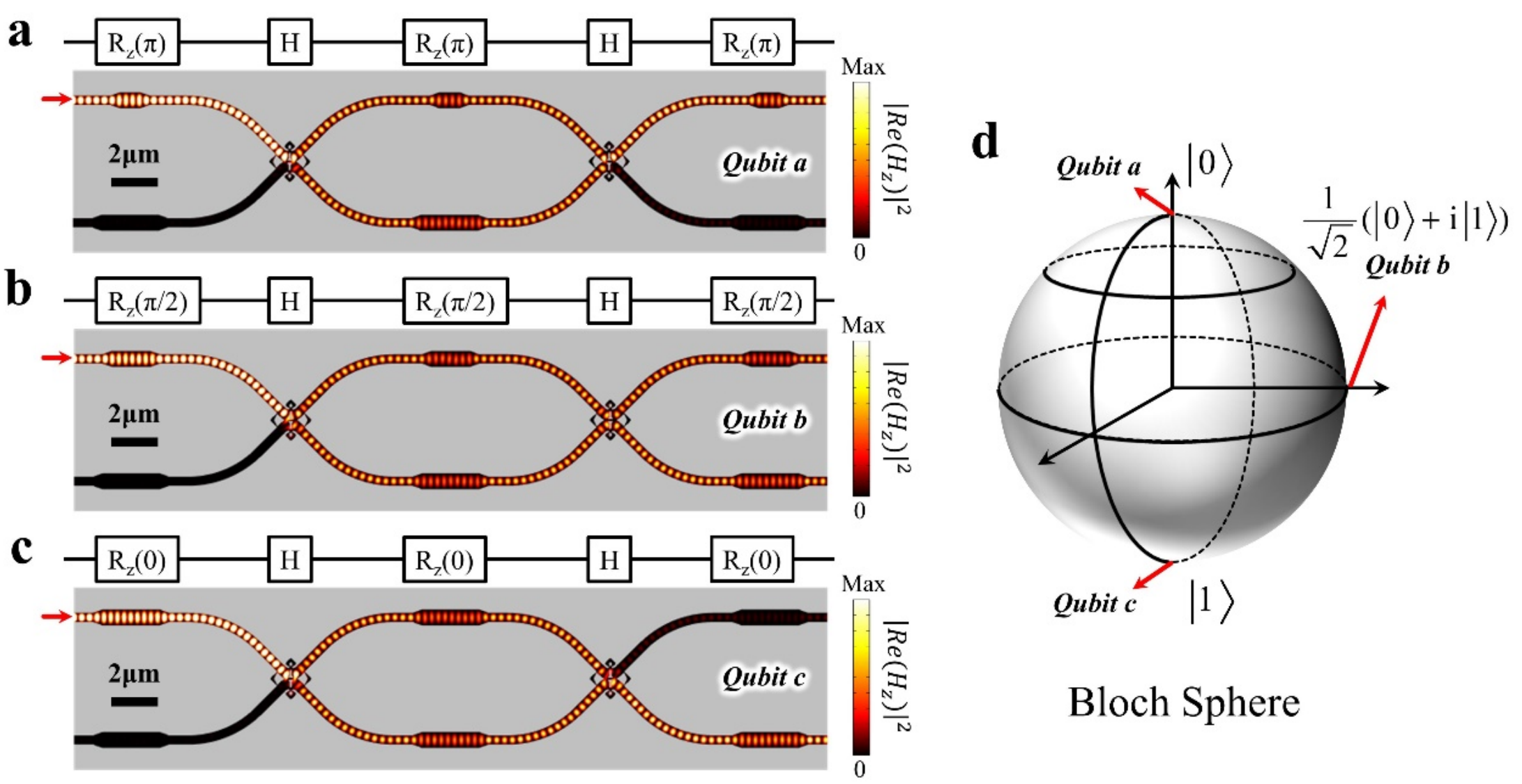

**Fig. S5. The simulation results of the arbitrary single qubit gate.** The phases of the phase z gates are set as (a) π, (b) π/2, (c) 0. (d) The Bloch Sphere of the single qubit gate R. The corresponding output qubits are marked on the Bloch Sphere.

Finally, let us explain the reason for the ultra-compact footprint of the cascaded R gate. In previous works, the quantum circuits are constructed by optical devices with large footprints and some connected waveguides on chips. The length of the waveguides is on the same order of magnitude with the size of optical devices. Due to the fabrication error, the phase accumulative effect causes an unstable phase difference between these waveguides. Thus, an additional heat electrode is indispensable to correct the phase difference between the path-encoded waveguides. The large optical devices and the connected waveguides with electrodes cover an enormous footprint with the scale of millimeters in traditional schemes. It brings about a great disadvantage for large-scale integrations. However, for our inverse-designed gate with a supercompact footprint (less than a wavelength), the length of the corresponding connected waveguide is about several micrometers. In this scale, the accumulative phase is so small that can be ignored. Thus, the R gate can be integrated into such a small area. Furthermore, the quantum circuits can be constructed by the R gates and the CNOT gates. In general, the size of quantum circuits can be effectively reduced from ~$10^7$μm$^2$ in the previous work to ~$10^3$μm$^2$ by using our inverse-designed quantum gates. The footprint is shrunk 4 orders of magnitude. The supercompact footprint of our devices indicates that more than $10^4$ quantum circuits can be integrated into the same area. For modern fabrication technology of the photonic chip, such a 2-qubit circuit (Ref. [5] of the main text) can be implemented based on the basic logic components of Hadamard and CNOT gates.

**S5. The imaginary parts for the density matrices of the bell states and the reconstructed process matrix of the CNOT gate.**

To completely characterize the CNOT gate and its generated bell states, we plot the imaginary parts for the density matrices of the bell states and the reconstructed process matrix of the CNOT gate, as shown in Fig. S6. It is seen that these imaginary parts can be limited to very small values (close to zero). Thus, we believe our inverse-designed CNOT gate possess a high performance.

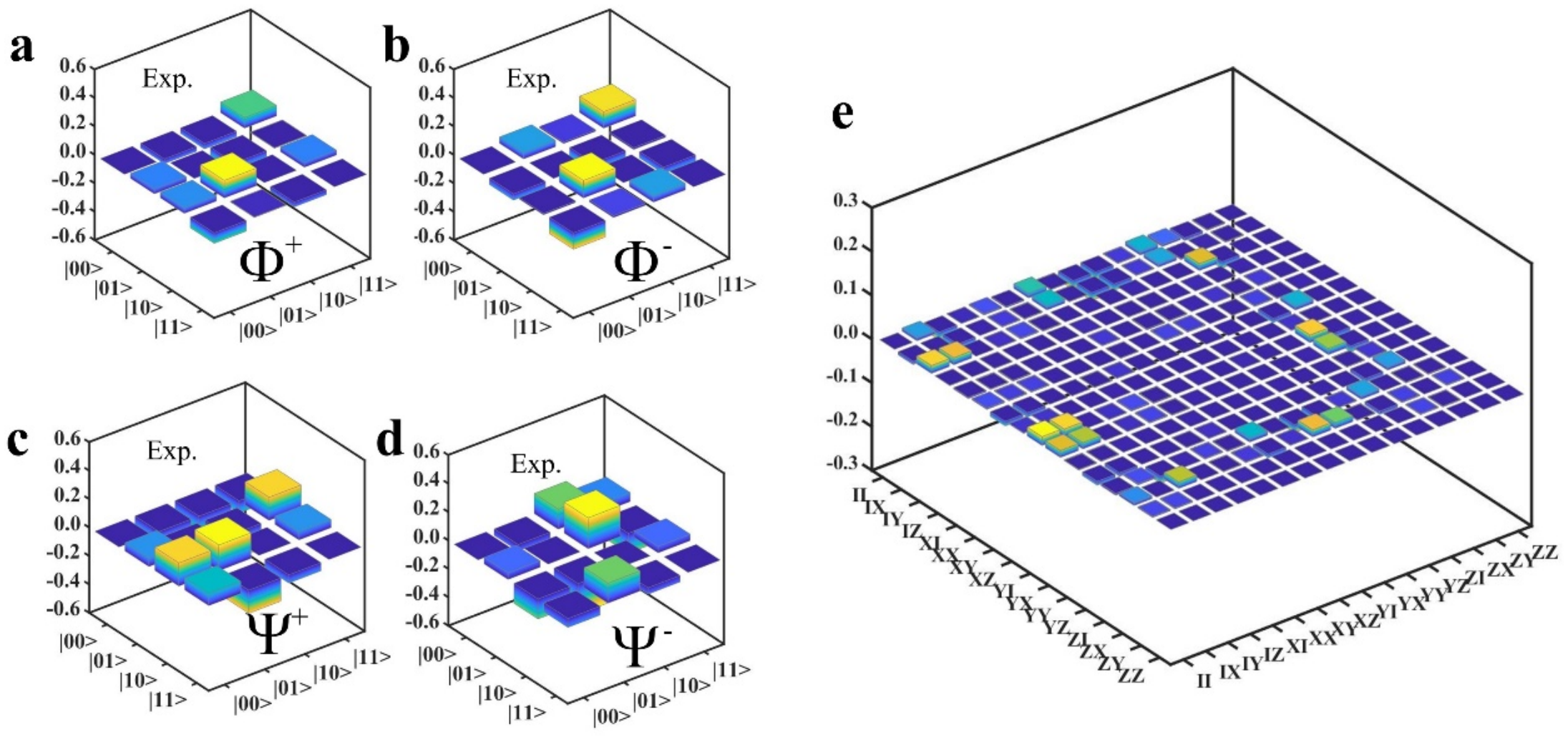

**Fig. S6. The experimental results of the imaginary parts for the reconstructed matrices.** The imaginary parts of the density matrices of the bell states (a) $\Phi^+$, (b) $\Phi^-$, (c) $\Psi^+$, and (d) $\Psi^-$. (e) The imaginary part of the reconstructed process matrix of the CNOT gate.

**S6. The overall size of the proposed quantum circuit.**

Here, we also show the overall size of the design layout for our proposed inverse-designed quantum circuit in Fig. S7a. This corresponds to the theoretical model shown in Fig. S7b. The overall size of the quantum circuit in Fig. 4 of the main text is about 100μm×25μm.

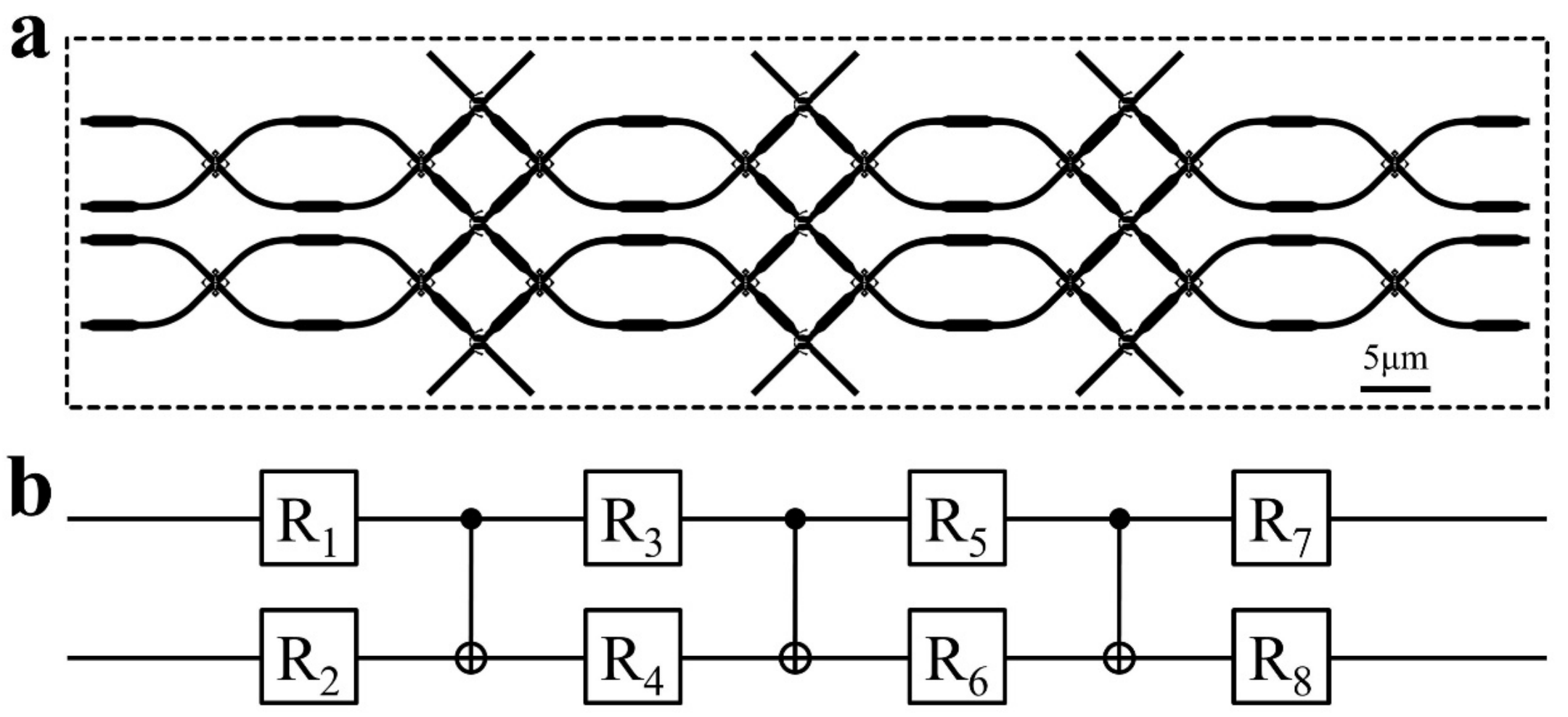

**Fig. S7. The 2-qubit quantum circuit by inverse-designed quantum gates.** (a) the design layout of the 2-qubit quantum circuit. (b) The corresponding theoretical model of the quantum circuit.

### S7. The simulation details of the optical transform matrix.

In Figs. 4b and 4d of the main text, we study the evolution process of the quantum state in the quantum circuit with the single photon exciting. The comparison figures of simulation and theory results are plotted in Figs. 4c and 4e of the main text. Next, we can also obtain the total optical matrix of the quantum circuit. Here, we mainly introduce the transform matrix method to obtain the theory results.

First, we build the transform matrices, $M_1$, $M_2$, and $M_3$, of these optical devices. Their expressions are

$$M_1 = \begin{pmatrix} r_1 & it_1 & & & & \\ it_1 & r_1 & & & & \\ & & r_1 & it_1 & & \\ & & it_1 & r_1 & & \\ & & & & r_1 & it_1 \\ & & & & it_1 & r_1 \end{pmatrix}, \tag{S3}$$

$$M_2 = \begin{pmatrix} 0 & & & & & \\ & r_2 & it_2 & & & \\ & it_2 & r_2 & & & \\ & & & r_2 & it_2 & \\ & & & it_2 & r_2 & \\ & & & & & 0 \end{pmatrix}, \tag{S4}$$

$$M_3 = \begin{pmatrix} 0 & & & & & \\ & 1 & & & & \\ & & e^{i\theta_1} & & & \\ & & & 1 & & \\ & & & & e^{i\theta_2} & \\ & & & & & 0 \end{pmatrix}, \tag{S5}$$

where $M_1$ is the transform matrix of the CNOT gate with $r_l$ ($t_l$) being $1/\sqrt{3}$ ($\sqrt{2/3}$). $M_2$ is the transform matrix of the Hadamard gate with $r_l$ ($t_l$) being $1/\sqrt{2}$. $M_3$ is the transform matrix of the phase z gate with $\theta_1$($\theta_2$) being a fixed phase. Next, we have the quantum state evolution equation:

$$|\varphi_{out}\rangle = M_3 M_2 M_3 M_2 M_3 M_1 M_3 M_2 |\varphi_{in}\rangle \tag{S6}$$

where $|\varphi_{out}\rangle$ ($|\varphi_{in}\rangle$) is the output (input) quantum state.

For simplicity, we set the phases of the phase z gates $\theta_1$ and $\theta_2$ as 0. The input states $|\varphi_{in}\rangle$ are set as $(0,1,0,0,0,0)^T$ and $(0,0,1,0,0,0)^T$, corresponding to Figs. 4b and 4d in the main text, respectively. By the calculation of Eq. (S5), the square of probability amplitude $|\varphi|^2$ of the output quantum state become $(0,1/6,1/6,1/3,1/3,0)^T$ and $(0,1/6,1/6,0,0,0)^T$, which are plotted in Figs. 4c and 4e (blue rectangle) in the main text. The theory results possess a nice consistency with the simulation results (marked as the red lines).

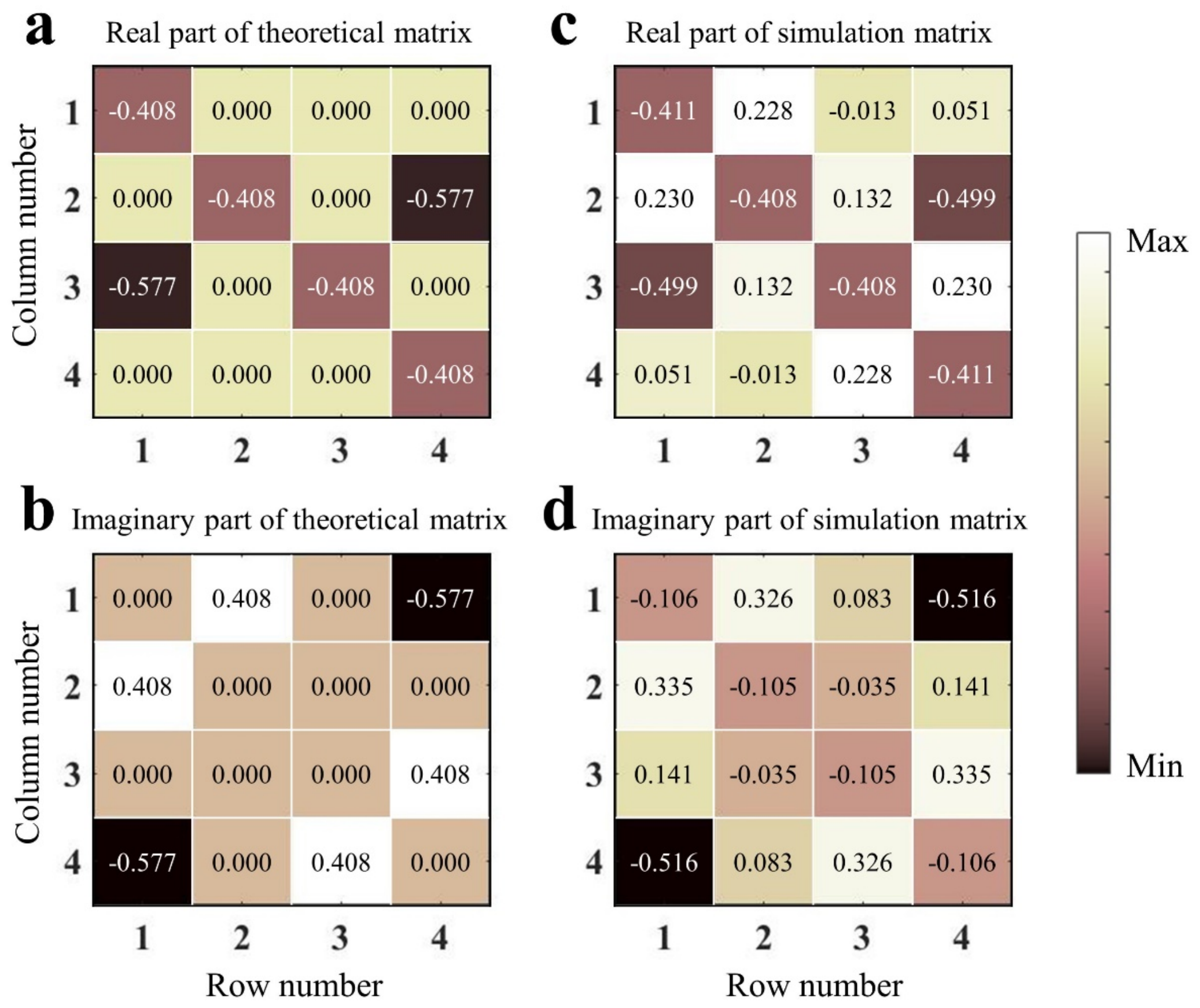

**Fig. S8. The total optical transform matrix of the quantum circuit.** a. The real part of the theoretical matrix. b. The imaginary part of the theoretical matrix. c. The real part of the simulation matrix. d. The imaginary part of the simulation matrix.

Moreover, we can also calculate the state transform matrices of the quantum circuit by theory and simulation methods. The theoretical transform matrix is obtained from Eq. (S5). The real and imaginary parts of the matrix are plotted in Figs. S8a and S8b. The simulation matrix elements are extracted from the quantum states from the output waveguides ($1^{st}$-$4^{th}$) in the full-wave simulation of the quantum circuit. The corresponding real and imaginary parts of the simulation matrix are plotted in Figs. S8c and S8d. Finally, we present the simulation results from our design, compared with the theoretical results, revealing an excellent agreement. It indicates that the circuit has high performance and can perform its function well.

## S8. The inverse-designed SWAP gate.

Here, we provide an implementation scheme for the specific 2-qubit gate, the SWAP gate. It consists of four Hadamard gates and three CNOT gates, as shown in Fig. S9a. In this way, a more compact footprint (~50μm$^2$) of the SWAP gate can be achieved. The encoding scheme of the SWAP gate is same to the CNOT gate in the main text.

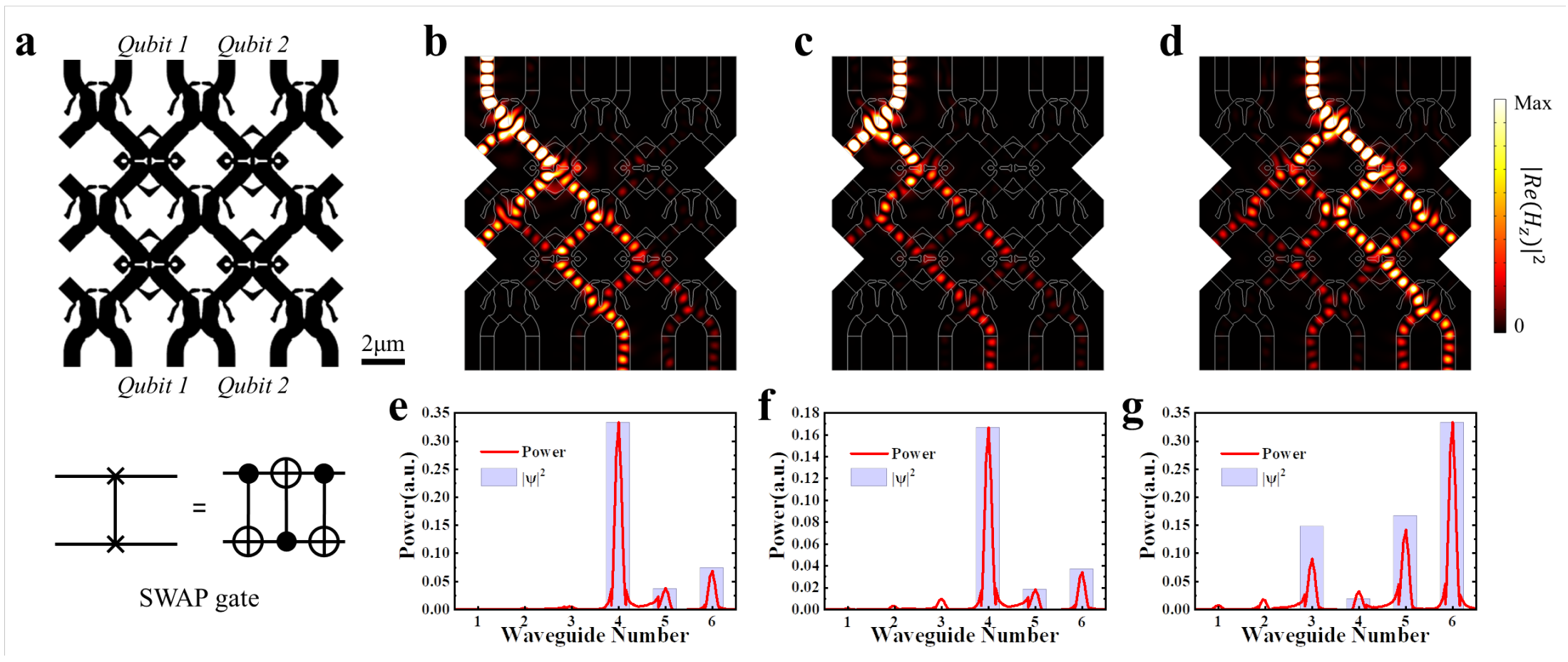

**Fig. S9. The inverse-designed SWAP gate. a**. The layout structure of the inverse-designed SWAP gate. **b-d.** The simulation results of the SWAP gate with the single photon state exciting. **e-g.** The comparison of the theory (blue rectangle) and simulation (red line) cases, corresponding to **b-d**, respectively.

And then, in order to demonstrate the function of the SWAP gate, we simulate the situation with a single photon exciting. The quantum state is injected in the 1$^{st}$, 2$^{nd}$, and 3$^{rd}$ waveguides, and the evolution processes are shown in Figs. S9b-S9d. Due to the symmetry of the SWAP gate, the evolution processes of the input quantum states exciting at the 4$^{th}$, 5$^{th}$, and 6$^{th}$ waveguides are same to the above cases. In order to test the function of the SWAP gate, we perform a numerical simulation of single-photon state evolution. Figs. S9b-S9d show squares of probability amplitudes ($|\varphi|^2$) with the single-photon state exciting from the 1$^{st}$ to 3$^{rd}$ waveguides, respectively. The corresponding $|\varphi|^2$ of output states from the 1$^{st}$ to 6$^{th}$ waveguides are marked as the red lines and shown in Figs. S9e-S9g.

On the other hand, for the theoretical description, we can also build the transform matrices. The quantum state evolution equation is

$$|\varphi_{out}\rangle = M_1 M_2 M_1 M_2 M_2 M_1 |\varphi_{in}\rangle \tag{S7}$$

where $\left|\varphi_{out}\right\rangle$ ($\left|\varphi_{in}\right\rangle$) is the output (input) quantum state of the SWAP gate. The input states $\left|\varphi_{in}\right\rangle$ are set as $(1,0,0,0,0,0)^{T}$, $(0,1,0,0,0,0)^{T}$, and $(0,0,1,0,0,0)^{T}$, corresponding to Figs. S9b, S9c, and S9d, respectively. By the calculation of Eq. (S6), the square of probability amplitude $|\varphi|^{2}$ of the output quantum state become $(0,0,0,0.333,0.037,0.074)^{T}$, $(0,0,0,0.167,0.019,0.037)^{T}$ and $(0,0,0.148,0.019,0.167,0.333)^{T}$, which are plotted in Figs. S9e-S9g (blue rectangle). The square of probability amplitude $|\varphi|^{2}$ of the quantum state is plotted with the theory (blue rectangle) and simulation (red line) case, as shown in Figs. S9e-S9g.

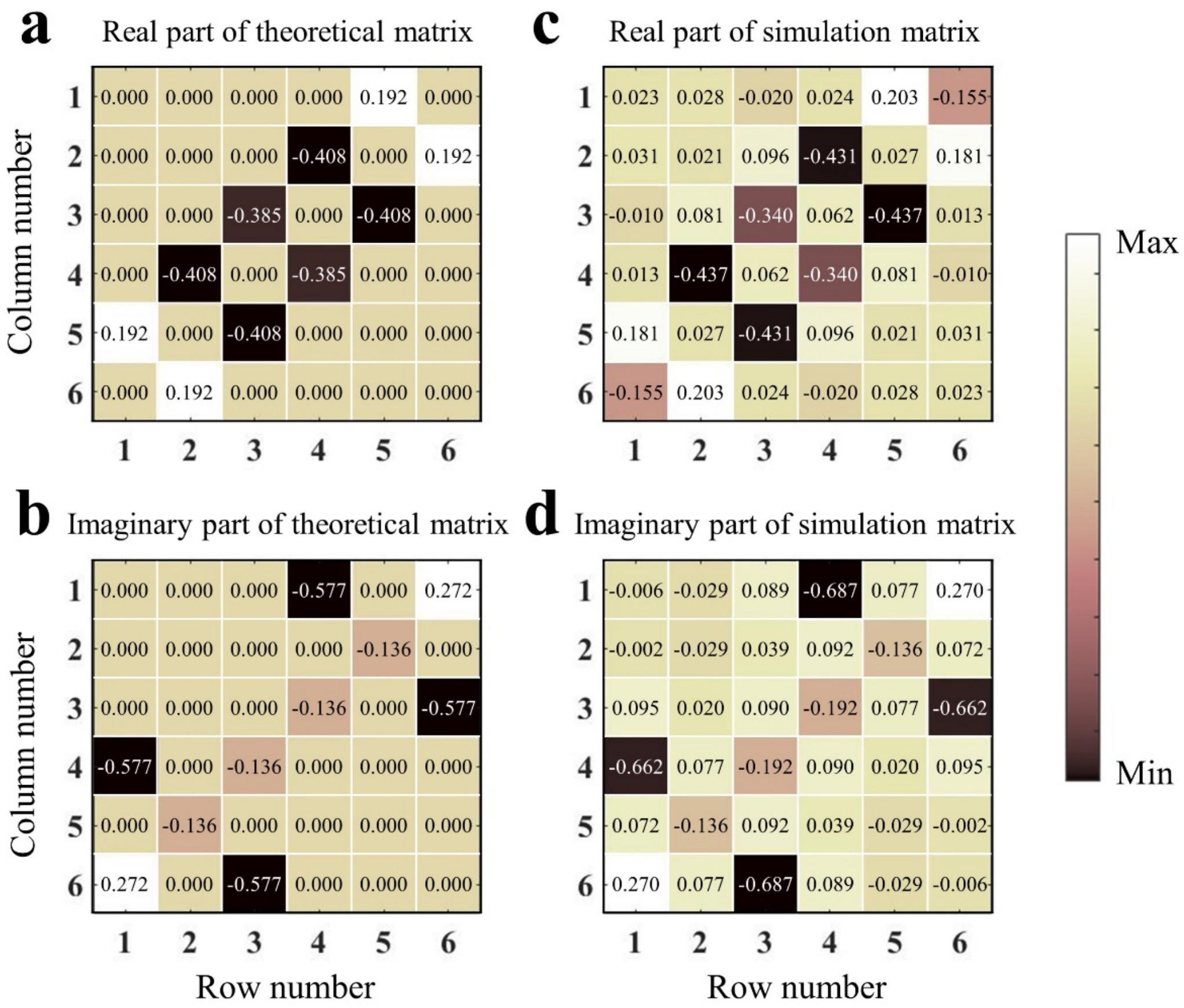

**Fig. S10. The total optical transform matrix of the SWAP gate.** a. The real part of the theoretical matrix. b. The imaginary part of the theoretical matrix. c. The real part of the simulation matrix. d. The imaginary part of the simulation matrix.

Moreover, we can also calculate the transform matrices of the SWAP gate by the theory and the simulation methods. The theoretical transform matrix is obtained from Eq. (S6). The real and imaginary parts of the matrix are plotted in Figs. S10a and S10b. The simulation matrix elements are extracted from the quantum states from the output waveguides ($1^{st}$-$6^{th}$) in the full-wave simulation of the SWAP gate. The corresponding real and imaginary parts of the simulation matrix are plotted in Figs. S10c and S10d. By comparison, we find that the theory results possess a nice consistency with the simulation results. It indicates that the circuit has a high performance and a low loss even though that is integrated into such a small footprint (~50$\mu m^2$). Thus, we believe our inverse-designed SWAP gate can perform its function well.